\documentclass[aps,prb,showpacs,tightenlines, twocolumn,secnumarabic,nofootinbib,nobibnotes,superscriptaddress]{revtex4-1}
\usepackage{amsmath,amssymb,amsfonts,bm}
\usepackage{graphicx}
\usepackage{epstopdf}
\usepackage{dcolumn}
\usepackage{mathrsfs}  
\usepackage{appendix}
\usepackage[colorlinks=true,linkcolor=blue,citecolor=blue, urlcolor=blue,bookmarks=false]{hyperref}

\begin{document}
		\title{Spatial anisotropy of Kondo screening cloud in a type-II Weyl semimetal}
	\author{Lu-Ji Wang}
	\author{Xing-Tai Hu}
	\affiliation{
		Department of Physics, Ningbo University, Ningbo 315211, China}
	\author{Lin Li} 
	\affiliation{
		College of Physics and Electronic Engineering, and Center for Computational Sciences,
		Sichuan Normal University, Chengdu 610068, China}
	\author{Dong-Hui Xu}
	\affiliation{
		Department of Physics, Hubei University, Wuhan 430062, China}
		\author{Jin-Hua Sun}
	\email{sunjinhua@nbu.edu.cn}
	\affiliation{
		Department of Physics, Ningbo University, Ningbo 315211, China}
	\author{Wei-Qiang Chen}
	\email{chenwq@sustc.edu.cn}
	\affiliation{
		Shenzhen Institute for Quantum Science and Engineering and Department of Physics,
		Southern University of Science and Technology, Shenzhen 518055, China}
	\affiliation{Center for Quantum Computing, Peng Cheng Laboratory, Shenzhen 518055, China}
	\begin{abstract}
    We theoretically study the Kondo screening of a spin-1/2 magnetic impurity in the bulk of a type-II Weyl semimetal (WSM) by use of the variational wave function method. 
    We consider a type-II WSM model with two Weyl nodes located on the $k_z$-axis, and the tilting of the Weyl cones are along the $k_x$ direction.   
    Due to co-existing electron and hole pockets, the density of states at the Fermi energy becomes finite, leading to a significant enhancement of Kondo effect. 
    Consequently, the magnetic impurity and the conduction electrons always form a bound state, this behavior is distinct from that in the type-I WSMs, where the bound state is only formed when the hybridization exceeds a critical value.         
    Meanwhile,     
    the spin-orbit coupling and unique geometry of the Fermi surface lead to strongly anisotropic Kondo screening cloud in coordinate space.
    The tilting terms break the rotational symmetry of the type-II WSM about the $k_z$-axis, but the system remains invariant under a combined transformation $\mathcal{T}R^{y}(\pi)$, where $\mathcal{T}$ is the time-reversal operation and $R^{y}(\pi)$ is the rotation about the $y$-axis by $\pi$. Largely modified diagonal and off-diagonal components of the spin-spin correlation function on three principal planes reflect this change in band symmetry. 
    Most saliently, the tilting terms trigger the emergence of non-zero off-diagonal components of spin-spin correlation function on the $x$-$z$ principal plane.

	\end{abstract}
	
	\maketitle

	\section{Introduction}
   
   As representatives of a new state of topological quantum matter, topological semimetals \cite{armitage2017} which host Dirac or Weyl fermions as low-energy excitations in the bulk have attracted much attention in recent years.  
   Three-dimensional (3D) Dirac semimetals have been realized experimentally in
   $\mathrm{ Na_3Bi}$\cite{liu2014} and $\mathrm{Cd_3As_2}$ materials,\cite{LiuZK2014,neupane2014} where the Dirac points are stabilized by the inversion ($\mathcal{P}$), time-reversal ($\mathcal{T}$) and crystalline symmetries. 
   If the $\mathcal{P}$ or/and $\mathcal{T}$ symmetry is broken, a transition towards the Weyl semimetal (WSM) phase takes place and each Dirac point splits into a pair of Weyl nodes. \cite{Wan2011,Burkov2011,Vazifeh2013}
   There has been tremendous interest in WSMs because a new TaAs family of WSMs was predicted theoretically \cite{Weng2015, Huang2015} and subsequently observed in experiments.\cite{Xu2015,Lv2015,Xu20152,Zhang2015,Yang2015,Wang2015,HuangXC2015,Lv2015724}
   The Weyl fermions in the TaAs family approximately respect the Lorentz symmetry. 
   However, the Weyl fermions realized in condensed matter physics are quasiparticles which can violate the Lorentz invariance, indicating that the Weyl cones in momentum space can be tilted. 
   
   The two-dimensional (2D) tilted anisotropic Dirac cones have
   been found in the 8-pmmn borophene\cite{Lopez2016} and in the organic
   semiconductor $\alpha$-(BEDT-TTF)$_2$I$_3$.\cite{Goerbig2008,Hirata2017} 
   In 3D systems, the band crossing points are more robust and generic than in 2D materials. 
   Type-II Dirac or Weyl fermions\cite{soluyanov2017, soluyanov2015, Xu2015prl} are obtained when Dirac or Weyl cones are tilted strongly in momentum space. In this case the electron and hole pockets co-exist with the Dirac or Weyl nodes. Type-II Weyl fermions are predicted and soon confirmed in $\text{WTe}_2$ and $\text{MoTe}_2$.\cite{soluyanov2015, Sunyan2015,Wang2016, Deng2016, Huang2016,Jiang2017} 
   Very strongly robust type-II Weyl nodes are predicted in 
   $\text{Ta}_3\text{S}_2$\cite{chang2016}, and observed in crystalline solid $\text{LaAlGe}$.\cite{XuSY2017} 
   Type-II WSMs show remarkable properties such as anisotropic chiral anomaly,\cite{soluyanov2015} unusual thermodynamic and optical responses in the presence of magnetic fields, \cite{O’Brien2016, Yang2016,Tchoumakov2016, Udagawa2016} and anomalous Hall effect.\cite{Ferreiros2017,saha2018}

  Kondo effect takes place when a magnetic impurity forms a singlet with the conduction electrons at the temperature lower than the Kondo temperature and has been widely studied by using various methods. \cite{Krishna1980,tsvelick1984,Andrei1984,Zhang1983,Coleman1984,read1983,Kuramoto1983,Gunnarsson1983,affleck1990}
  In systems with isotropic Dirac cones, the magnetic impurity problem falls into the category of the pseudogap Kondo problem,\cite{Gonzalez1998,Fritz2004,Vojta2004} and has been constantly studied\cite{Chang2015,Ulloa2016,Shun2016,Zheng2016} in recent years following the discoveries of various novel host systems in condensed matter physics. 
  There exists a critical value of hybridization for the impurity and conduction electrons to form a bound state.\cite{Feng2010, shirakawa2014}
  On the other hand, the spin-orbit couplings in many of the systems lead to very rich features in the spin-spin correlation function between the magnetic impurity and the conduction electrons.\cite{Feng2010,Liu2009} 
%
  
%
  In the type-II WSM, the topology is compeletely unchanged by the tilting terms in comparison with the conventional type-I WSM.
  However, the type-II WSM has Fermi surfaces instead of Weyl nodes and thus gives rise to a finite density of states(DOS)\cite{Udagawa2016} at the Fermi energy.  
  The binding energy and the spatial spin-spin correlation of a magnetic impurity can reflect these changes in host materials. 
  Hence the remarkable electronic structures of a type-II WSM are expected to largely modify the behavior of a magnetic impurity embedded in the bulk. 
  Indeed we find that the binding energy and the spin-spin correlation between the magnetic impurity and conduction electrons show distinctions in comparison with their counterparts in a type-I WSM,\cite{Jinhua2015} especially in the emergence of non-zero off-diagonal correlation functions on the $x$-$z$ coordinate plane.

  In this paper, we systematically investigate the binding energy and spatial spin-spin correlation function between a spin-1/2 magnetic impurity and the conduction electrons in a type-II WSM.
  We use the variational wave function method to perform the calculations.   
  The variational method we apply has been used to study the ground state of the Kondo problem in normal metals,\cite{Gunnarsson1983,Varma1976} antiferromagnet,\cite{Aji2008} 2D helical metals,\cite{Feng2010} and various novel topological materials.\cite{Jinhua2015, Ma2018, Lv2019, Jinhua2018, deng2018}
 The paper is organized as follows. We present the model Hamiltonian, dispersion as well as the electron and hole pockets at the Fermi level in Sec. \ref{Sec:Hamiltonian}. 
  In Sec. \ref{Sec:selfconsist}, we apply the variational method to study the binding energy and present the differences caused by the tilting terms. 
  In Sec. \ref{Sec:sscorr}, we calculate the spin-spin correlation between the magnetic impurity and the conduction electrons in a type-II WSM on three principal planes in coordinate space and analyze the results.
  Finally, the discussions and conclusions are given in
  Sec. \ref{Sec:conclusion}.

	\section{Hamiltonian}\label{Sec:Hamiltonian}
    
    We use the Anderson impurity model to study the Kondo screening of a spin-1/2 magnetic impurity in a type-II WSM, 
    the total Hamiltonian is given by 
    \begin{equation}\label{Eq:totalH}
    \begin{aligned}
    H=H_0 + H_d + H_V.
    \end{aligned}
    \end{equation}
    $H_0$ is the kinetic energy term, $H_d$ describes the magnetic impurity part, and $H_V$ is the hybridization between the local impurity and the conduction electrons. 
    The low-energy effective Hamiltonian of a type-II WSM in momentum space is given by  
	\begin{equation}\label{Eq:H0}
	H_0=\sum_{\mathbf{k}}\Psi_{\mathbf{k}}^\dagger\left[ h_0(\mathbf{k})-\mu \right] \Psi_{\mathbf{k}}, 
	\end{equation}
	with
	\begin{equation}\label{Eq:h0detail}
	\begin{aligned}
	h_0(\mathbf{k})&=t^\prime \tau_z(\sigma_xk_x+\sigma_yk_y)+t\tau_z\sigma_0k_z+\tau_x\sigma_0M_{\mathbf{k}} \\
	& +b\tau_0\sigma_z+(a_{tilt}k_x+{\xi}k_x^2/2)\tau_0\sigma_0 .
	\end{aligned}
	\end{equation}
	$h_0(\mathbf{k})$ is obtained by expanding the lattice model Hamiltonian\cite{O’Brien2016} (with lattice constant $a_0=1$) of a type-II WSM around the Weyl nodes. The Fermi energy is fixed as $\mu=0$ throughout this work.
	The basis vectors are given by  $\Psi_{\mathbf{k}}= \{a_{\mathbf{k}\uparrow},  a_{\mathbf{k}\downarrow}, b_{\mathbf{k}\uparrow}, b_{\mathbf{k}\downarrow}\}^T$,  where $a_{\mathbf{k}s}^\dagger$ ($b_{\mathbf{k}s}$) creates (annihilates) an electron with spin-$s$ ($s=\uparrow, \downarrow$) on the $a$ ($b$) orbit. 
	$\sigma_{\alpha}$ and $\tau_{\alpha}$ ($\alpha=x,y,z$) are the spin and orbital Pauli matrices. 
	In principle $t$ and $t^\prime$ can be different, but here we fix $t=t^\prime$ and set them as the energy unit,  
	in order to eliminate extra anisotropy caused.
    $M_\mathbf{k}$ is obtained by expanding the term $m_0-\sum_{\alpha} \cos k_\alpha $ around the Weyl nodes, where the Dirac mass is $m_0 = 3t$. Notably $h_0(\mathbf{k})$ differs from the conventional type-I WSM Hamiltonian\cite{Vazifeh2013} by additional $a_{tilt}$ and $\xi$ terms. 
    Moreover, in order to stop the electron and hole pockets from spreading over the entire Brillouin zone, 
    the term $\tau_y\sigma_0 \sin k_z$ is replaced by $\tau_z\sigma_0\sin k_z$\cite{O’Brien2016}. 
    In the original type-I WSM Hamiltonian given in Ref. \cite{Vazifeh2013} 
    in the absence of $b$, $a_{tilt}$ and $\xi$, $H_0$ describes a Dirac semimetal with degenerate Dirac points located at $\mathbf{k}=0$. 
    A nonzero $b$ breaks the time-reversal symmetry, and a type-I WSM emerges with a pair of Weyl nodes at $(0, 0, \pm b/t)$ on the $k_z$-axis. 
    The transition from a type-I to type-II WSM takes place when $a_{tilt}$ increases sufficiently that the Weyl cones are strongly tilted along the $k_x$ direction leading to coexisting electron and hole states on the Fermi
    surface. $\xi$ further breaks the symmetry between the electron and hole pockets around each Weyl node. 
    
    \begin{figure}[t]
	\begin{center}
		\includegraphics[scale=0.5, bb=60 20 400 430]{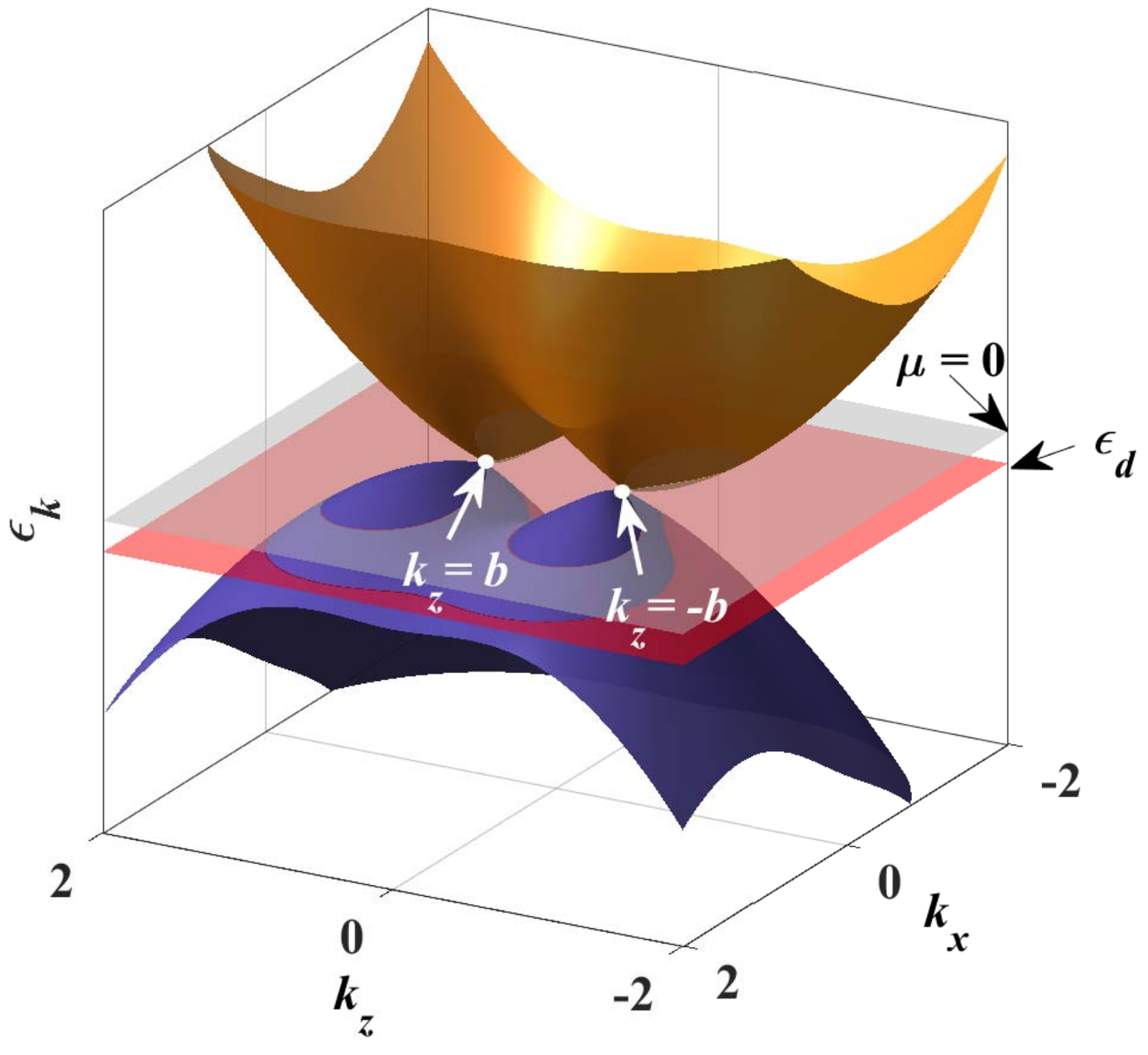}
	\end{center}
	\caption{(Color online) Schematic of the band structure of a type-II WSM for $k_y=0$. $b$ determines the distance between the pair of Weyl nodes in a type-I WSM, $a_{tilt}$ tilts the Weyl cones along the $k_x$ axis generating a pair of electron and hole pockets on the Fermi surface, and $\xi$ breaks the symmetry between the electron and the hole pockets. The Fermi energy is fixed as $\mu=0$, and energy level of the magnetic impurity is $\epsilon_d<\mu$, so for a large enough $U$ the impurity is always singly occupied.    
	} \label{0_band}
\end{figure}

    The single particle eigenenergy is given by 
	\begin{equation}\label{Eq:ek}
	\begin{aligned}
	\epsilon_\mathbf{k}=\pm\sqrt{\eta_\mathbf{k}\pm2\nu_\mathbf{k}}+n_\mathbf{k},
	\end{aligned}
	\end{equation}
    where $\nu_\mathbf{k}\equiv \sqrt{t^2k_z^2\left[b^2+t^{\prime2}(k_x^2+k_y^2)\right]+b^2M_\mathbf{k}^2}$, $\eta_\mathbf{k}\equiv b^2+t'^{2}(k_x^2+k_y^2)+t^2k_z^2+M_\mathbf{k}^2$ and $n_\mathbf{k}\equiv  a_{tilt}k_x+{\xi}k_x^2/2$. 
    $H_0$ in its diagonal basis reads 
	\begin{equation}
	\begin{aligned}
	H_0 = \sum_{\mathbf{k}}\Psi_{\mathbf{k}}^\dagger  h_0(\mathbf{k})   \Psi_{\mathbf{k}} =  \sum_{\mathbf{k}i} \epsilon_{\mathbf{k}i} \gamma_{\mathbf{k}i}^\dagger \gamma_{\mathbf{k}i}, \ (i=1,2,3,4).  \\
	\end{aligned}
	\end{equation}
    The relation between the eigenstates $\gamma_{\mathbf{k}i}^\dagger$ and $\gamma_{\mathbf{k}i}$ and the original electron creation and annihilation operators are given in the appendix.

    \begin{figure}[t]
	\begin{center}
		\includegraphics[scale=0.8, bb=160 290 460 580]{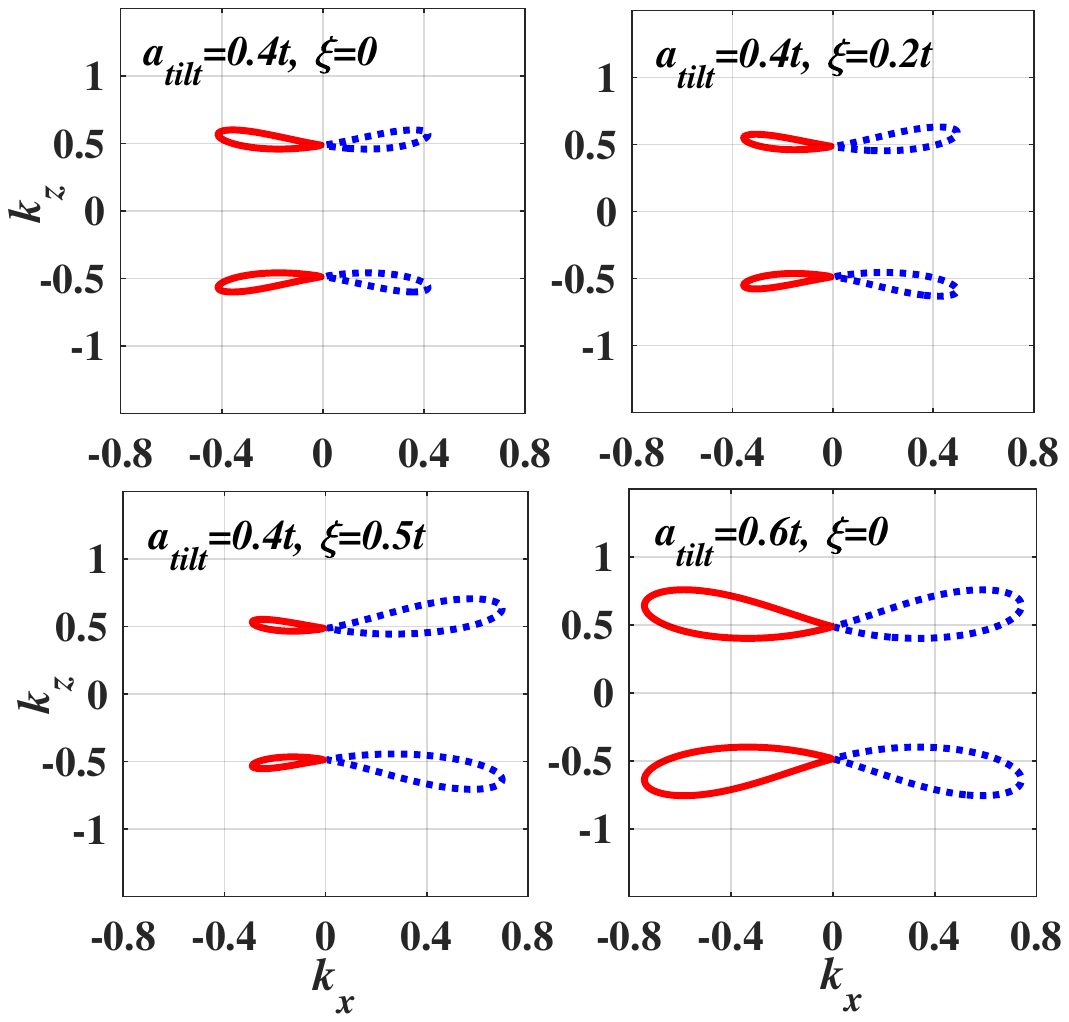}
	\end{center}
	\caption{(Color online) 
		The electron and hole pockets on the $k_x$-$k_z$ plane for $b=0.5t$, with different combinations of $a_{tilt}$ and $\xi$. } \label{0_pocket}
\end{figure}

	The localized state is described by
	\begin{equation}\label{Eq:Hamil_Impu}
	\begin{aligned}
	H_{d} &=\epsilon _d\sum _{s=\uparrow,\downarrow }d_{s }^{\dagger}d_{s }+\text{U}d_{\uparrow }^{\dagger}d_{\uparrow }d_{\downarrow
	}^{\dagger}d_{\downarrow }.
	\end{aligned}
	\end{equation}
	$d_{s}^\dagger$ and $d_{s}$ are the creation and annihilation operators of the
	spin-$s$ ($s=\uparrow, \downarrow$) state on the impurity site.
	$\epsilon_d$ is the impurity energy level, and $U$ is the on-site Coulomb repulsion.  
	
	Finally, the hybridization term between the localized state and the electron spins in the type-II WSM is given by
	\begin{equation}\label{Eq:Hv}
	\begin{aligned}
     H_{V}&=\sum _{\mathbf{k},s=\uparrow,\downarrow }V_\mathbf{k}\left[(a_{\mathbf{k}s }^\dagger+b_{\mathbf{k}s }^\dagger)d_s +H.c.\right].
	\end{aligned}
	\end{equation}
	Here $V_\mathbf{k} \equiv V \Theta \left[\Gamma - |\epsilon(\mathbf{k})|\right]$, where $\Theta(x)$ is a step function, which is $1$ for $x>0$ and $0$ for $x<0$. $\Gamma$ is the energy cut-off and is chosen as a large enough value, such that the low-energy physics is expected to be insensitive to the value of $\Gamma$. 
    The impurity is equally coupled to the a, b orbits, and to the spin-up and -down states.  
    In the diagonal basis of the type-II WSM, the hybridization part $H_V$ reads 
    \begin{equation}\label{Eq:Hv}
    \begin{aligned}
    H_{V}&=\sum _{\mathbf{k}i }V_\mathbf{k}\left( \gamma_{\mathbf{k}i}^\dagger d_{\mathbf{k}i} +H.c.\right). 
    \end{aligned}
    \end{equation}
    The $\mathbf{k}$-dependent impurity operators are connected to the original ones through transformation
	\begin{equation}
	\begin{aligned}
    d_{\mathbf{k}i}^{\dagger }&=\left[ \left(\Phi_{1i}+\Phi_{3i}\right)d_{\uparrow}^{\dagger} +\left(\Phi_{2i}+\Phi_{4i}\right)d_{\downarrow}^{\dagger} \right]\\
    & = \chi_{i1}(\mathbf{k}) \ d_{\uparrow}^\dagger + \chi_{i2}(\mathbf{k}) \ d_{\downarrow}^\dagger, 
	\end{aligned}
	\end{equation}
	where $i=1,2,3,4$ are the eeband indices, and the definition of $\Phi_{ij}$ is given in the appendix. 
	
	In Fig. \ref{0_band} we show the schematic of the dispersion of a type-II WSM for $k_y=0$.   
	The two Weyl nodes are located on $k_z = \pm b/t$, and relatively large $a_{tilt}$ term generates a pair of electron and hole pockets around each Weyl node.    
	The $\xi$ term breaks the symmetry between the electron and hole pockets. 
	Throughout this work, the Fermi energy is fixed as $\mu=0$, and the magnetic impurity energy level is $\epsilon_d<\mu$. 
	For large enough $U$ the impurity site shall be always singly occupied. 
	
	In Fig. \ref{0_pocket} we plot the electron and hole pockets for $b=0.5t$. The electron and hole pockets only emerge when the tilting term $a_{tilt}$ becomes large enough.\cite{O’Brien2016}
	We can see that while $\xi=0$, for both $a=0.4t$ and $0.6t$, the electron and hole pockets are symmetric. Finite $\xi=0.2t$ breaks the symmetry between the pockets around each Weyl node when $a_{tilt}=0.4t$. 
	As $\xi$ increases, the asymmetry between the pockets becomes more significant. 
	The tilting terms modify the DOS at the Fermi energy and also break the rotational symmetry about the $z$-axis of the type-II WSM model Hamiltonian. Hence the binding energy and the spatial Kondo screening cloud are expected to be distinct from those in a conventional type-I WSM.

	\section{The self-consistent calculation }\label{Sec:selfconsist}
	In order to investigate the eigenstate property, we utilize a trial wavefunction approach. 
	The Coulomb repulsion $U$ is assumed to be large enough, and $\epsilon_d$ is below the Fermi energy, such that the impurity site is always singly occupied with a local moment.
	First, we may assume $H_V=0$, which is the simplest case that the magnetic impurity and the host material is completely decoupled from each other. The ground state of $H_0$ is given by
	\begin{equation}
	\begin{aligned}
    |\Psi_0 \rangle =\prod _{\mathbf{k}\in \Omega, i} \ \gamma _{\mathbf{k}i}{}^{\dagger}|0\rangle. 
	\end{aligned}
	\end{equation}
    $i$ is the band index, and the product runs over all states within the Fermi sea $\Omega$.
    If we consider about a singly occupied impurity, and ignore the energy given by the hybridization, then the total energy of the system is just the sum of bare impurity energy and total energy of the WSM,
	\begin{equation}
	\begin{aligned}
	E_0=\epsilon _d +\sum _{\mathbf{k}\in\Omega,i}\epsilon_{\mathbf{k}i}. 
	\end{aligned}
	\end{equation}
	
	If the hybridization is taken into account, the trial wave function for the ground state shall be
	\begin{equation}\label{Eq:newwavef}
	\begin{aligned}
	|\Psi \rangle =\left(a_0+\sum _{\mathbf{k}\in\Omega,i}a_{\mathbf{k}i}d_{\mathbf{k}i}^{\dagger}\gamma_{\mathbf{k}i}\right)|\Psi_0\rangle.
	\end{aligned} 
	\end{equation}
	$a_{0}$, $a_{\mathbf{k}i}$ are all numbers and they are the variational parameters to be determined through self-consistent calculations.  
	The energy of total Hamiltonian in the variational state $|\Psi \rangle $ shall be
	\begin{equation}\label{Eq:energy}
	\begin{aligned}
	E=\frac{\langle\Psi |H|\Psi\rangle}{\langle\Psi|\Psi\rangle}.
	\end{aligned}
	\end{equation}
	We can obtain $\langle\Psi|\Psi\rangle=a_0^2+\sum_{\mathbf{k}\in\Omega,i}a_{\mathbf{k}i}^2(|\chi_{i1}(\mathbf{k})|^2+|\chi_{i2}(\mathbf{k})|^2)=1$ according to the wavefunction normalization condition.

		\begin{figure}[t]
	\begin{center}
		\includegraphics[scale=0.45, bb=80 60 550 530]{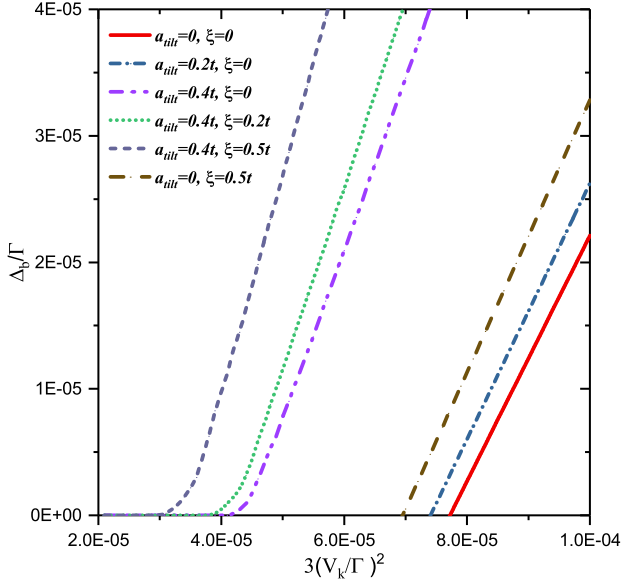}
	\end{center}
	\caption{(Color online). The self-consistent results of the binding energy with $b=0.5t$ for various combinations of $a_{tilt}$ and $\xi$. $\mu=0$ and $\epsilon_d=7.5\times 10^{-5} \Gamma$, where $\Gamma$ is the energy cutoff. There exist a critical $V_k$ to form a positive binding energy when the DOS at the Fermi energy is zero when \{$a_{tilt}=0$, $\xi=0$\}, \{$a_{tilt}=0.2t$, $\xi=0$\} or \{$a_{tilt}=0$, $\xi=0.5t$\}. Otherwise when the electron and hole pockets are formed as shown in Fig. \ref{0_pocket}, the DOS becomes nonzero at the Fermi energy, so the binding energy is always positive although the magnitude is very small when the value of $V_k$ is small.} \label{bindingenergy}
\end{figure}

\begin{figure*}[htb!]
	\begin{center}
		\includegraphics[scale=0.68, bb=280 80 550 350]{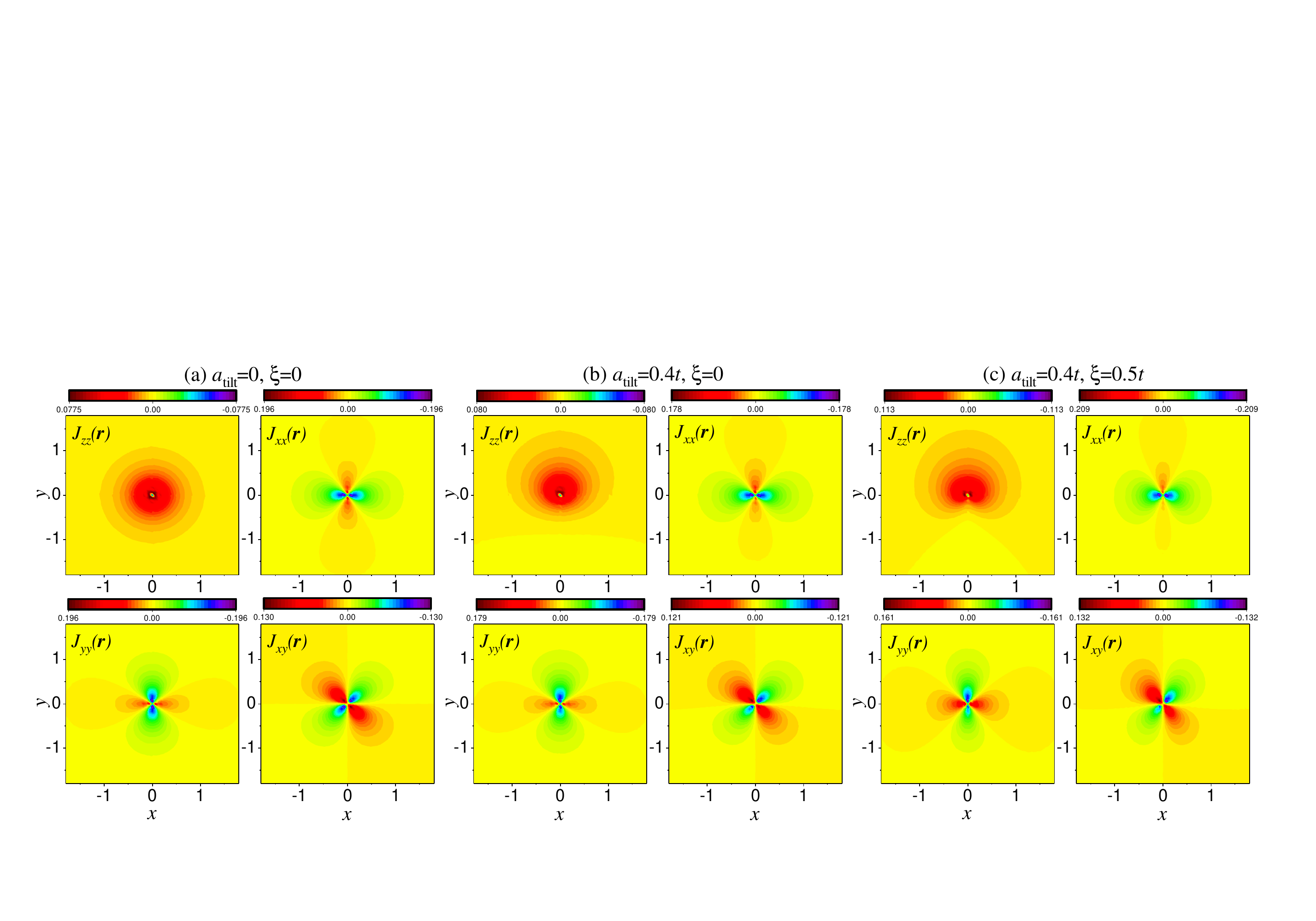}
	\end{center}
	\caption{(Color online). Terms of the spin-spin correlation $J_{uv}$ ($u,v=x,y,z$) on the $x$-$y$ coordinate space. All the other off-diagonal terms not shown is zero on the $x$-$y$ plane. In all the plots $b=0.5t$ and the tilting terms are (a) $a_{tilt}=0$, $\xi=0$, (b) $a_{tilt}=0.4t$, $\xi=0$, (c) $a_{tilt}=0.4t$, $\xi=0.5t$. } \label{1_xy}
\end{figure*}
	
	Then the total energy of the type-II Weyl system with a magnetic impurity in the trial state $|\Psi \rangle$ writes
	
	\begin{equation}\label{Eq:totalE}
	\begin{aligned}
	E=&\sum_{\mathbf{k}\in\Omega,i}[ (E_0-\epsilon_{\mathbf{k}i}+\mu)(|\chi_{i1}(\mathbf{k})|^2+|\chi_{i2}(\mathbf{k})|^2)a_{\mathbf{k}i}^2+\\ &2V_\mathbf{k}a_0a_{\mathbf{k}i}(|\chi_{i1}(\mathbf{k})|^2+|\chi_{i2}(\mathbf{k})|^2)+(\epsilon_{\mathbf{k}i}-\mu)a_0^2]/
	[a_0^2+\\
	&\sum_{\mathbf{k}\in\Omega,i}a_{\mathbf{k}i}^2(|\chi_{i1}(\mathbf{k})|^2+|\chi_{i2}(\mathbf{k})|^2)].
	\end{aligned}
	\end{equation}
	
	The variational principle requires that $\partial E/\partial a_0=\partial E/\partial a_{\mathbf{k}}=0$, leading to two equations below:
	\begin{equation}\label{Eq:threeab}
	\begin{gathered}
	\left(E-\sum_{\mathbf{k}\in\Omega,i}\epsilon_{\mathbf{k}i}\right)a_0 = \sum_{\mathbf{k}\in\Omega,i}V_\mathbf{k}a_{\mathbf{k}i}\left(|\chi_{i1}(\mathbf{k})|^2+|\chi_{i2}(\mathbf{k})|^2\right),\\
	\left(E-E_0+\epsilon_{\mathbf{k}i}\right)a_{\mathbf{k}i} = V_\mathbf{k}a_0.\\
	\end{gathered}
	\end{equation}
	
	We then obtain the self-consistent equation
	\begin{equation}\label{Eq:selfConsis}
	\begin{aligned}
	\epsilon_d  - \Delta_b = \sum_{\mathbf{k}\in\Omega,i} \frac{V_\mathbf{k}^2(|\chi_{i1}(\mathbf{k})|^2+|\chi_{i2}(\mathbf{k})|^2)}{\epsilon_{\mathbf{k}i}- \Delta_b },  \\
	\end{aligned}
	\end{equation}
	$\Delta_b=E_0-E$ is the binding energy. If $\Delta_b>0$, the hybridized state has lower energy and is more stable than the bare state.
	$\Delta_b$ can be obtained by numerically solving the self-consistent equation given in Eq. \ref{Eq:selfConsis}. $a_0$ and $a_{\mathbf{k}i}$ for each value of $\mathbf{k}$ and $i$ can be calculated according to the relations
	\begin{equation}\label{Eq:a0ak}
	\begin{gathered}
	a_0^2 + \sum_{\mathbf{k}\in\Omega,i}a_{\mathbf{k}i}^2(|\chi_{i1}(\mathbf{k})|^2+|\chi_{i2}(\mathbf{k})|^2)=1,\\
	a_{\mathbf{k}i} = \frac{V_\mathbf{k}}{\epsilon_{\mathbf{k}i}-\Delta_b}a_0.\\
	\end{gathered}
	\end{equation}
	
    In Fig. \ref{bindingenergy} we present the self-consistent results of $\Delta_b$ as a function of $V_k/\Gamma$ for various combinations of $a_{tilt}$ and $\xi$. The results are obtained by numerically solving Eq. \ref{Eq:selfConsis}. 
    Here we fix the value of $b=0.5t$, and $\Gamma$ is the energy cut-off.   
    When $a_{tilt}=0$ and $\xi=0$, $H_0$ describes a type-I WSM, such that the DOS at the Fermi energy vanishes. In this case, the magnetic impurity problem falls into the category of pseudogap Kondo problem.\cite{Gonzalez1998,Fritz2004,Vojta2004} The magnetic impurity and the conduction electron spins form a bound state only if the hybridization is stronger than a critical value.\cite{Jinhua2015}   
    If we slightly tilt the Weyl nodes ($a_{tilt}=0.2t$, $\xi=0$ or $a_{tilt}=0$, $\xi=0.5t$), the electron and hole pockets are not formed on the Fermi surface, so the DOS at the Fermi energy is still zero. Similar to the case of a type-I WSM, $\Delta_b$ is positive only if $V_k$ is larger than a critical value, but the values of $\Delta_b$ slightly increase for the same hybridization strength, indicating that for the tilted system the bound state is more easily formed although the DOS at the Fermi energy is still zero.   
    If we go on to increase the tilting term to $a_{tilt}=0.4t$, as given in Fig. \ref{0_pocket}, a pair of electron and hole pockets emerge around each Weyl node, leading to a finite DOS at the Fermi energy. We can see that for $a_{tilt}=0.4t$, $\Delta_b$ for small $V_k$ is close to zero but becomes positive. It means that for any small but positive values of $V_k$ the impurity and the host material always form a bound state. If a nonzero value of $\xi$ is added, then the electron and hole pockets become asymmetric, leading to a larger value of DOS at the Fermi energy. Hence for these cases the binding energy becomes larger than the symmetric case when $\xi$ is zero.

	\section{Spin-spin correlation}\label{Sec:sscorr}
	
	In this section, we study the spin-spin correlation between the magnetic impurity and the conduction electrons in type-II WSMs. 
	The spin operators of the magnetic impurity and conduction electrons in type-II WSMs are defined as $\mathbf{S_d}=\frac{1}{2} d^{\dagger}\vec{\sigma} d$, $\mathbf{S_a}=\frac{1}{2} a^{\dagger}\vec{\sigma} a$ and $\mathbf{S_b}=\frac{1}{2} b^{\dagger}\vec{\sigma} b$. $d=\{d_{\uparrow}, d_{\downarrow}\}^T$, $a=\{a_{\uparrow}, a_{\downarrow}\}^T$, $b=\{b_{\uparrow}, b_{\downarrow}\}^T$ are the annihilation operators on impurity site and on the two orbits in the type-II WSM, respectively.    
	Without loss of generality, we choose the position of magnetic impurity as $\mathbf{r}=0$. Consequently, in momentum space, the impurity is equally coupled to each band, and the hybridization $V_k$ is in fact independent of $k$. 
	
	Both the $a$ and $b$ orbits contribute to the spin-spin correlation between the magnetic impurity and the conduction electron located at $\mathbf{r}$. The correlation function consists of two parts,     
	$J_{uv}(\mathbf{r})= \langle S_{a}^u (\mathbf{r})S_d^v(0)+S_{b}^u (\mathbf{r})S_d^v(0)\rangle=J_{uv}^a(\mathbf{r})+J_{uv}^b(\mathbf{r})$. The first term is the $a$-orbital contribution while the second term is $b$-orbital.
	Here $u,v=x,y,z$, and $\langle \cdots \rangle$ denotes the ground state average. 

	The magnitude of the binding energy $\Delta_b$ depends directly on the DOS at the Fermi energy. 
	In a Dirac semimetal or in a type-I WSM, the DOS vanishes at the Dirac points or Weyl nodes, so there exists a threshold of the hybridization strength for a positive $\Delta_b$.
	However, if one tunes $\mu$ away from the Dirac points or the Weyl nodes, the DOS at the Fermi energy becomes finite. $\Delta_b$ always has a positive solution, that the localized state and the conduction electrons form bound states for arbitrarily small $V_k$. 
	On the other hand, once the bound states are formed, the spatial spin-spin correlation functions are not much affected by the choice of $\mu$ except for the magnitude. 
	In this present paper, the spin-spin correlation function is evaluated for $\mu=0$.  
	The diagonal and the off-diagonal terms of the spin-spin correlation in coordinate space are given by
    Eq. \ref{Eq:sscorr} in the appendix. 
    For a finite value of $\mu$, the spatial patterns of the various components of the spin-spin correlation are expected to be qualitatively the same.

  \begin{figure*}[htb!]
  	\begin{center}
  		\includegraphics[scale=0.68, bb=280 80 550 350]{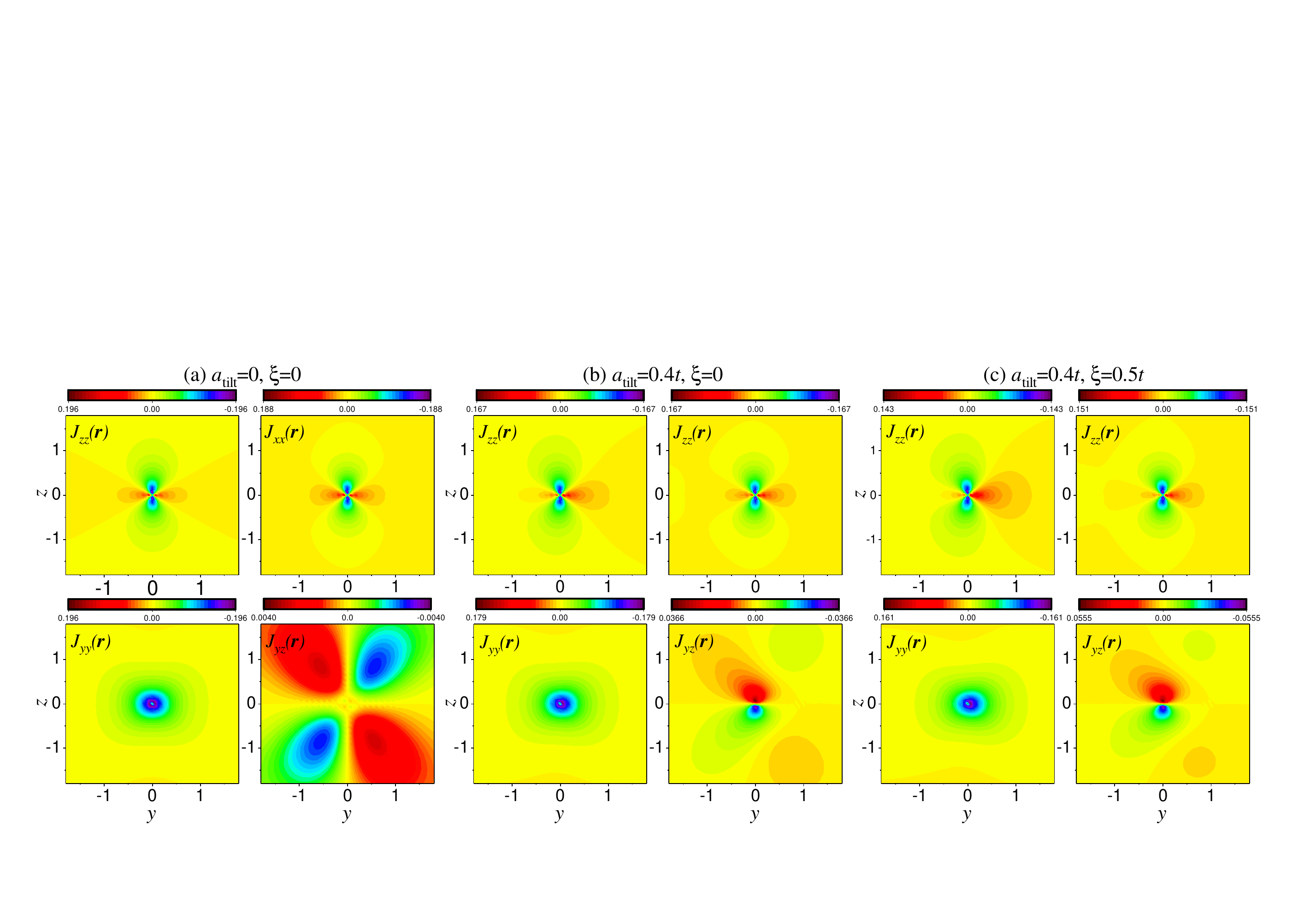}
  	\end{center}
  	\caption{(Color online). Terms of the spin-spin correlation $J_{uv}$ ($u,v=x,y,z$) on the $y$-$z$ plane for $b=0.5t$. All the other off-diagonal terms not shown is zero on the $y$-$z$ plane. (a) $a_{tilt}=0$, $\xi=0$, (b) $a_{tilt}=0.4t$, $\xi=0$, (c) $a_{tilt}=0.4t$, $\xi=0.5t$. } \label{2_yz}
  \end{figure*}

\begin{figure}[htb!]
	\begin{center}
		\includegraphics[scale=0.45, bb=80 80 550 780]{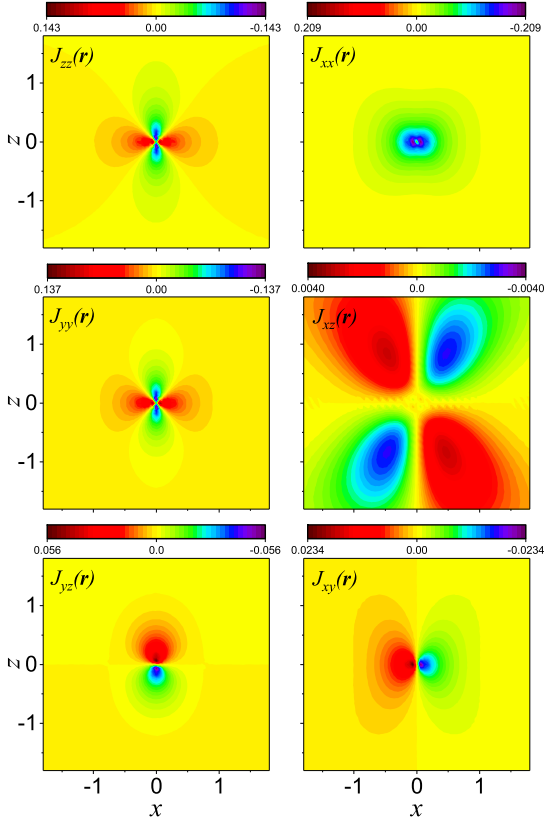}
	\end{center}
	\caption{(Color online). Terms of the spin-spin correlation $J_{uv}(\mathbf{r})$ ($u,v=x,y,z$) on the $x$-$z$ coordinate space for $b=0.5t$, $a_{tilt}=0.4t$ and $\xi=0.5t$. Due to the tilting terms the off-diagonal terms $J_{yz}(\mathbf{r})$ and $J_{xy}(\mathbf{r})$ shown nonzero values on the $x$-$z$ plane.}
	\label{3_xz}
\end{figure}

In Fig. \ref{1_xy} - Fig. \ref{3_xz} we show the results of the spin-spin correlation between the local magnetic impurity and the conduction electrons on the $x$-$y$, $y$-$z$ and $x$-$z$ plane in the coordinate space.
We fix $b=0.5t$, and three typical combinations of tilting terms are: (1) $a_{tilt}=\xi=0$ representing a type-I WSM, (2) $a_{tilt}=0.4t$ and $\xi=0$ with symmetric electron and hole pockets and (3) $a_{tilt}=0.4t$ and $\xi=0.5t$ representing a type-II WSM with asymmetric electron and hole pockets. 

In the first case, the time-reversal symmetry is broken, but the system preserves the rotational symmetry about the $z$-axis, so we have $J_{uv}(\mathbf{r}) = J_{u'v'}(\mathbf{r'})$  if $u^{\prime} = R^z(\beta)u$, $v^{\prime} = R^z(\beta)v$, $\mathbf{r^{\prime}} = R^z(\beta)(\mathbf{r})$, 
where $R^z(\beta)$ is a rotation operator about the $z$-axis. 

As the $a_{tilt}$ and $\xi$ terms become finite the rotational symmetry about the $z$-axis is broken, but one can easily demonstrate that the Hamiltonian is still invariant under a combined operation $\mathcal{T}R^y(\pi)$, where $\mathcal{T}$ is the time-reversal operation and $R^{y}(\pi)$ is a rotation of angle $\pi$ about the $y$ direction. 
Under the transformation $\mathcal{T}R^{y}(\pi)$ we have 
\begin{equation}
\begin{aligned}
\{x,y,z\}\rightarrow\{-x, y, -z\}, \\
\{k_x,k_y,k_z\}\rightarrow\{k_x, -k_y, k_z\},\\
\{s_x,s_y,s_z\}\rightarrow\{s_x,-s_y,s_z\}.
\end{aligned}\label{Eq:TRypi}
\end{equation}  

Large enough $a_{tilt}$ generates a pair of electron and hole pockets around each Weyl node, and a non-zero $\xi$ triggers the asymmetry between the electron and hole pockets as plotted in Fig. \ref{0_pocket}. 
The change in the band structure and DOS due to the $a_{tilt}$ and $\xi$ terms naturally leads to the modifications in the spin-spin correlation between the magnetic impurity and the conduction electron spins. 
In fact, the binding energy $\Delta_b$ shall take different values while the model parameters change. 
However, we may fix the value of $\Delta_b$ in the spin-spin correlation calculations in order to mainly concentrate on the spatial patterns. The parameter we use in this section is $V_k = 0.1t$ and $\Delta_b=0.1t$. 
The length unit is chosen as $1/k_d$ where $k_d$ is the momentum cut-off.
The values of $\mathcal{A}_{mn}(\mathbf{r})$ given in Eq. \ref{Eq:21} are complex numbers in general, so natually the off-diagonal terms $J_{uv}(\mathbf{r})\neq J_{vu}(\mathbf{r})$, ($u,v=x,y,z$). 
However, we find that $J_{uv}(\mathbf{r})$ and $J_{vu}(\mathbf{r})$ shows similar patterns with same symmetry property on the three principal planes. Hence we only plot the components $J_{xz}(\mathbf{r})$, $J_{yz}(\mathbf{r})$ and $J_{xy}(\mathbf{r})$ in the maintext, and others are discussed and plotted in the appendix.    
A positive (negative) value of the diagonal component indicates the ferromagnetic (antiferromagnetic) correlation between the impurity spin and the conduction electron spin. 


In Fig. \ref{1_xy} we show the results of the diagonal and off-diagonal terms of the spin-spin correlation between the magnetic impurity and the conduction electrons on the $y$-$z$ plane in coordinate space. 
In Fig. \ref{1_xy} (a) the tilting terms vanish ($a_{tilt}=\xi=0$), so the Hamiltonian describes a Type-I WSM with two Weyl nodes located at $\pm b/t$ on the $k_z$-axis. 
$b$ breaks the time-reversal symmetry, but the system still preserves the rotational symmetry about the $z$-axis.  
Hence in Fig. \ref{1_xy} (a) $J_{zz}(\mathbf{r})$ has rotational symmetry on the $x$-$y$ plane, and the correlation is antiferromagnetic nearby the magnetic impurity, and oscillates as $|\mathbf{r}|$ increases. 
The other two diagonal terms have the relation $J_{xx}(\mathbf{r})=J_{yy}(R^z(\pi/2)\mathbf{r})$, and both are ferromagnetic along one real space axis while are antiferromagnetic along the other axis. 
Among the off-diagonal terms, only $J_{xy}(\mathbf{r})$ is nonzero. 
By carefully examining the results we find that the terms $J_{xz}^a(\mathbf{r})=-J_{xz}^b(\mathbf{r})$, $J_{yz}^a(\mathbf{r})=-J_{yz}^b(\mathbf{r})$, so finally the off-diagonal components $J_{xz}(\mathbf{r})$ and $J_{yz}(\mathbf{r})$ vanish on the $x$-$y$ plane. 
According to the transformation given in Eq. \ref{Eq:TRypi} $J_{xy}(x,y)=-J_{xy}(-x,y)$, and if $x=0$ the off-diagonal term $J_{xy}(\mathbf{r})$ is always zero. 
This is valid even if the tilting terms are added, as given in Fig. \ref{1_xy} (b) and (c) since the system is still invariant under  $\mathcal{T}R^{y}(\pi)$. 
When the $a_{tilt}$ term becomes finite as shown in Fig. \ref{1_xy} (b), all the four terms of the spin-spin correlation function lose the rotational symmetry of $\pi$ about the $z$ direction. We can see that all the three diagonal terms are tilted along the $y$-axis, and this change is most obvious in $J_{zz}(\mathbf{r})$. The magnitude of the off-diagonal term $J_{xy}(\mathbf{r})$ also becomes asymmetric with respect to the $x$-axis.   
If the term $\xi$ is also imposed as is shown in Fig. \ref{1_xy} (c), the rotational symmetry is further broken. The magnitude of spin-spin correlation shows much stronger anisotropy.   

 Plotted in Fig. \ref{2_yz} are the components of the spin-spin correlation on the $y$-$z$ principal plane. Among the off-diagonal terms, only $J_{yz}(\mathbf{r})$ is nonzero.
 $J_{xz}(\mathbf{r})$ and $J_{xy}(\mathbf{r})$ vanish because the $a$-orbital and $b$-orbital contributions cancel with each other.  
  In Fig. \ref{2_yz} (a) we show the spin-spin correlation for the type-I WSM. The system preserves the rotational symmetry about the $z$-axis.
  Consequently, all the three diagonal terms show $J_{uu}(\mathbf{r})=J_{uu}(R^z(\pi)\mathbf{r})$ ($u=x,y,z$). 
  Moreover, due to the $TR^{y}(\pi)$ symmetry, the diagonal terms also exhibit the property $J_{uu}(y,z)=J_{uu}(y,-z)$. As to the off-diagonal term we have $J_{yz}(\mathbf{r})=J_{yz}(R^y(\pi)\mathbf{r})$ and $J_{yz}(y,z)=-J_{yz}(y,-z)$.
  With finite $a_{tilt}$ and $\xi$ as in Fig. \ref{2_yz} (b) and (c), the rotational symmetry is broken, and the WSM is only invariant under the operation $TR^{y}(\pi)$. 
 In the presence of a finite $a_{tilt}$ as in Fig. \ref{2_yz} (b), we can see that the rotational symmetry of $\pi$ of spin-spin correlations is broken. However, the diagonal terms have the property $J_{zz}(y,z)=J_{zz}(y,-z)$, $J_{xx}(y,z)=J_{xx}(y,-z)$ and $J_{yy}(y,z) = J_{yy}(y,-z)$ due to the transformation given in Eq. \ref{Eq:TRypi}. 
 The off-diagonal term is $J_{yz}(y,z)=-J_{yz}(y,-z)$. 
 Even if the tilting term $\xi$ is added, the system is still invariant under the combined $TR^{y}(\pi)$ transformation, such that diagonal terms are symmetric about the $z$-axis while the off-diagonal term is $J_{yz}(y,z)=-J_{yz}(y,-z)$.

 In Fig. \ref{3_xz} we show the spin-spin correlation function on the $x$-$z$ coordinate space for $b=0.5t$, $a_{tilt}=0.4t$ and $\xi=0.5t$. 
 For the case $b=0.5t$ in the absence of tilting terms, the system has rotational symmetry about the $z$-axis, and is also invariant under $TR^{y}(\pi)$. In this case, the results on the $x$-$z$ plane have a direct relation with those on $y$-$z$ plane, that $J_{zz}(x,z) = J_{zz}(y,z)$, $J_{xx}(x,z) = J_{yy}(y,z)$ and $J_{yy}(x,z)=J_{xx}(y,z)$.  
 Among the off-diagonal terms, only $J_{xz}(\mathbf{r})$ is nonzero and it is related to the correlation on the $y$-$z$ plane by $J_{xz}(x,z) = J_{yz}(y,z)$. Hence when $a_{tilt}=\xi=0$, we can relate all the non-zero components of spin-spin correlation on the $x$-$z$ plane with those on the $y$-$z$ plane. 
  
 Very interestingly, the tilting term $a_{tilt}=0.4t$ triggers non-zero off-diagonal components $J_{yz}(\mathbf{r})$ and $J_{xy}(\mathbf{r})$ on the $x$-$z$ plane. If a nonzero $\xi$ is added, the spatial pattern of the correlations are slightly modified, but the symmetry properties remain the same so we only show the results of $b=0.5t$, $a_{tilt}=0.4t$ and $\xi=0.5t$ in Fig. \ref{3_xz}. 
 Once the tilting terms become finite, the rotational symmetry about the $z$-axis is broken, but the system is still invariant under the transformation $TR^{y}(\pi)$. 
 Hence the diagonal terms $J_{xx}(\mathbf{r})$, $J_{yy}(\mathbf{r})$ and $J_{zz}(\mathbf{r})$ show inversion symmetry on the $x$-$z$ plane, which can be given as $J_{uu}(x,z) = J_{uu}(-x,-z)$. 
 The off-diagonal term $J_{xz}(\mathbf{r})$ also shows the same inversion symmetry, while $J_{yz}(\mathbf{r})=-J_{yz}(\mathbf{r})$ and $J_{xy}(\mathbf{r})=-J_{xy}(\mathbf{-r})$ since the spin operator $s_y\rightarrow -s_y$ under the operation $TR^{y}(\pi)$ as given in Eq. \ref{Eq:TRypi}.

%
%

\section{conclusions}\label{Sec:conclusion}
	In summary, we have utilized the variational wave function method to investigate the binding energy and the spatial anisotropy of the Kondo screening cloud in a type-II WSM. 
	The type-II WSM is defined by a continuous four-band model Hamiltonian, with a pair of Weyl nodes located on the $k_z$-axis. In the presence of tilting terms, the Weyl cones are tilted along the $k_x$ direction forming pairs of electron and hole pockets. 
    The DOS becomes finite at the Fermi energy, so the Kondo effect is significantly enhanced. The bound state is always favored by the magnetic impurity and the type-II WSMs. This behavior is distinct from that in a type-I WSM, where the bound state is only formed if $V_\mathbf{k} > V_c$,\cite{Jinhua2015} where $V_c$ is a threshold of hybridization strength. 
    The spatial spin-spin correlation function shows very strong anisotropy due to the spin-orbit coupling and the unique band structure of the type-II system. 
    The topology of the type-II WSM is the same as the type-I WSM, but the geometry of the bands and the DOS become distinct.
    The tilting terms $a_{tilt}$ and $\xi$ break the rotational symmetry about the $z$-direction. However, the type-II WSM model Hamiltonian remains invariant under $\mathcal{T}R^y(\pi)$. 
    Our spin-spin correlation results reflect these changes in the host materials. 
    All the non-zero components of the spin-spin correlation function on the three principal planes are largely modified by the tilting terms. 
    The most significant changes are the emergence of several non-zero off-diagonal correlation functions in type-II WSMs on the $x$-$z$ coordinate plane. 
    It has been theoretically suggested that the topology and the form of Fermi surface of a type-II WSM are very sensitive to pressure, strain and elastic deformation.\cite{soluyanov2015,zubkov2018} 
    This offers as an opportunity to tune the Kondo effect in various regimes in the type-II WSMs. The type-II WSM also shows unique Fermi arc surface states,\cite{Zheng2018} and we will address the issue of magnetic impurity in the novel surface states in future work.

%

	\section{Acknowledgments}
	J.H.S. acknowledges financial support from the NSFC (Grant No. 11604166), Zhejiang Provincial Natural Science Foundation of
	China (Grant No. LY19A040003) and K.C.Wong Magna Fund in Ningbo University.
	L.L. is supported by the NSFC (under Grant No. 11604138).  
	D.H.X. is supported by the NSFC (under Grant No. 11704106). W.Q.C acknowledges financial support from National Key Research and Development Program of China (No. 2016YFA0300300) and NSFC (No. 11674151).
	
	\newpage
\bibliographystyle{apsrev4-1}
\bibliography{ref}

\begin{thebibliography}{64}%
\makeatletter
\providecommand \@ifxundefined [1]{%
 \@ifx{#1\undefined}
}%
\providecommand \@ifnum [1]{%
 \ifnum #1\expandafter \@firstoftwo
 \else \expandafter \@secondoftwo
 \fi
}%
\providecommand \@ifx [1]{%
 \ifx #1\expandafter \@firstoftwo
 \else \expandafter \@secondoftwo
 \fi
}%
\providecommand \natexlab [1]{#1}%
\providecommand \enquote  [1]{``#1''}%
\providecommand \bibnamefont  [1]{#1}%
\providecommand \bibfnamefont [1]{#1}%
\providecommand \citenamefont [1]{#1}%
\providecommand \href@noop [0]{\@secondoftwo}%
\providecommand \href [0]{\begingroup \@sanitize@url \@href}%
\providecommand \@href[1]{\@@startlink{#1}\@@href}%
\providecommand \@@href[1]{\endgroup#1\@@endlink}%
\providecommand \@sanitize@url [0]{\catcode `\\12\catcode `\$12\catcode
  `\&12\catcode `\#12\catcode `\^12\catcode `\_12\catcode `\%12\relax}%
\providecommand \@@startlink[1]{}%
\providecommand \@@endlink[0]{}%
\providecommand \url  [0]{\begingroup\@sanitize@url \@url }%
\providecommand \@url [1]{\endgroup\@href {#1}{\urlprefix }}%
\providecommand \urlprefix  [0]{URL }%
\providecommand \Eprint [0]{\href }%
\providecommand \doibase [0]{http://dx.doi.org/}%
\providecommand \selectlanguage [0]{\@gobble}%
\providecommand \bibinfo  [0]{\@secondoftwo}%
\providecommand \bibfield  [0]{\@secondoftwo}%
\providecommand \translation [1]{[#1]}%
\providecommand \BibitemOpen [0]{}%
\providecommand \bibitemStop [0]{}%
\providecommand \bibitemNoStop [0]{.\EOS\space}%
\providecommand \EOS [0]{\spacefactor3000\relax}%
\providecommand \BibitemShut  [1]{\csname bibitem#1\endcsname}%
\let\auto@bib@innerbib\@empty
\bibitem [{\citenamefont {Armitage}\ \emph {et~al.}(2018)\citenamefont
  {Armitage}, \citenamefont {Mele},\ and\ \citenamefont
  {Vishwanath}}]{armitage2017}%
  \BibitemOpen
  \bibfield  {author} {\bibinfo {author} {\bibfnamefont {N.~P.}\ \bibnamefont
  {Armitage}}, \bibinfo {author} {\bibfnamefont {E.~J.}\ \bibnamefont {Mele}},
  \ and\ \bibinfo {author} {\bibfnamefont {A.}~\bibnamefont {Vishwanath}},\
  }\href {\doibase 10.1103/RevModPhys.90.015001} {\bibfield  {journal}
  {\bibinfo  {journal} {Rev. Mod. Phys.}\ }\textbf {\bibinfo {volume} {90}},\
  \bibinfo {pages} {015001} (\bibinfo {year} {2018})}\BibitemShut {NoStop}%
\bibitem [{\citenamefont {Liu}\ \emph {et~al.}(2014{\natexlab{a}})\citenamefont
  {Liu}, \citenamefont {Zhou}, \citenamefont {Zhang}, \citenamefont {Wang},
  \citenamefont {Weng}, \citenamefont {Prabhakaran}, \citenamefont {Mo},
  \citenamefont {Shen}, \citenamefont {Fang}, \citenamefont {Dai} \emph
  {et~al.}}]{liu2014}%
  \BibitemOpen
  \bibfield  {author} {\bibinfo {author} {\bibfnamefont {Z.}~\bibnamefont
  {Liu}}, \bibinfo {author} {\bibfnamefont {B.}~\bibnamefont {Zhou}}, \bibinfo
  {author} {\bibfnamefont {Y.}~\bibnamefont {Zhang}}, \bibinfo {author}
  {\bibfnamefont {Z.}~\bibnamefont {Wang}}, \bibinfo {author} {\bibfnamefont
  {H.}~\bibnamefont {Weng}}, \bibinfo {author} {\bibfnamefont {D.}~\bibnamefont
  {Prabhakaran}}, \bibinfo {author} {\bibfnamefont {S.-K.}\ \bibnamefont {Mo}},
  \bibinfo {author} {\bibfnamefont {Z.}~\bibnamefont {Shen}}, \bibinfo {author}
  {\bibfnamefont {Z.}~\bibnamefont {Fang}}, \bibinfo {author} {\bibfnamefont
  {X.}~\bibnamefont {Dai}},  \emph {et~al.},\ }\href
  {http://science.sciencemag.org/content/343/6173/864} {\bibfield  {journal}
  {\bibinfo  {journal} {Science}\ }\textbf {\bibinfo {volume} {343}},\ \bibinfo
  {pages} {864} (\bibinfo {year} {2014}{\natexlab{a}})}\BibitemShut {NoStop}%
\bibitem [{\citenamefont {Liu}\ \emph {et~al.}(2014{\natexlab{b}})\citenamefont
  {Liu}, \citenamefont {Jiang}, \citenamefont {Zhou}, \citenamefont {Wang},
  \citenamefont {Zhang}, \citenamefont {Weng}, \citenamefont {Prabhakaran},
  \citenamefont {Mo}, \citenamefont {Peng}, \citenamefont {Dudin},
  \citenamefont {Kim}, \citenamefont {Hoesch}, \citenamefont {Fang},
  \citenamefont {Dai}, \citenamefont {Shen}, \citenamefont {Feng},
  \citenamefont {Hussain},\ and\ \citenamefont {Chen}}]{LiuZK2014}%
  \BibitemOpen
  \bibfield  {author} {\bibinfo {author} {\bibfnamefont {Z.}~\bibnamefont
  {Liu}}, \bibinfo {author} {\bibfnamefont {J.}~\bibnamefont {Jiang}}, \bibinfo
  {author} {\bibfnamefont {B.}~\bibnamefont {Zhou}}, \bibinfo {author}
  {\bibfnamefont {Z.}~\bibnamefont {Wang}}, \bibinfo {author} {\bibfnamefont
  {Y.}~\bibnamefont {Zhang}}, \bibinfo {author} {\bibfnamefont
  {H.}~\bibnamefont {Weng}}, \bibinfo {author} {\bibfnamefont {D.}~\bibnamefont
  {Prabhakaran}}, \bibinfo {author} {\bibfnamefont {S.-K.}\ \bibnamefont {Mo}},
  \bibinfo {author} {\bibfnamefont {H.}~\bibnamefont {Peng}}, \bibinfo {author}
  {\bibfnamefont {P.}~\bibnamefont {Dudin}}, \bibinfo {author} {\bibfnamefont
  {T.}~\bibnamefont {Kim}}, \bibinfo {author} {\bibfnamefont {M.}~\bibnamefont
  {Hoesch}}, \bibinfo {author} {\bibfnamefont {Z.}~\bibnamefont {Fang}},
  \bibinfo {author} {\bibfnamefont {X.}~\bibnamefont {Dai}}, \bibinfo {author}
  {\bibfnamefont {Z.}~\bibnamefont {Shen}}, \bibinfo {author} {\bibfnamefont
  {D.}~\bibnamefont {Feng}}, \bibinfo {author} {\bibfnamefont {Z.}~\bibnamefont
  {Hussain}}, \ and\ \bibinfo {author} {\bibfnamefont {Y.}~\bibnamefont
  {Chen}},\ }\href {\doibase 10.1038/nmat3990} {\bibfield  {journal} {\bibinfo
  {journal} {Nature Materials}\ }\textbf {\bibinfo {volume} {13}},\ \bibinfo
  {pages} {677} (\bibinfo {year} {2014}{\natexlab{b}})}\BibitemShut {NoStop}%
\bibitem [{\citenamefont {Neupane}\ \emph {et~al.}(2014)\citenamefont
  {Neupane}, \citenamefont {Xu}, \citenamefont {Sankar}, \citenamefont
  {Alidoust}, \citenamefont {Bian}, \citenamefont {Liu}, \citenamefont
  {Belopolski}, \citenamefont {Chang}, \citenamefont {Jeng}, \citenamefont
  {Lin} \emph {et~al.}}]{neupane2014}%
  \BibitemOpen
  \bibfield  {author} {\bibinfo {author} {\bibfnamefont {M.}~\bibnamefont
  {Neupane}}, \bibinfo {author} {\bibfnamefont {S.-Y.}\ \bibnamefont {Xu}},
  \bibinfo {author} {\bibfnamefont {R.}~\bibnamefont {Sankar}}, \bibinfo
  {author} {\bibfnamefont {N.}~\bibnamefont {Alidoust}}, \bibinfo {author}
  {\bibfnamefont {G.}~\bibnamefont {Bian}}, \bibinfo {author} {\bibfnamefont
  {C.}~\bibnamefont {Liu}}, \bibinfo {author} {\bibfnamefont {I.}~\bibnamefont
  {Belopolski}}, \bibinfo {author} {\bibfnamefont {T.-R.}\ \bibnamefont
  {Chang}}, \bibinfo {author} {\bibfnamefont {H.-T.}\ \bibnamefont {Jeng}},
  \bibinfo {author} {\bibfnamefont {H.}~\bibnamefont {Lin}},  \emph {et~al.},\
  }\href {https://www.nature.com/articles/ncomms4786} {\bibfield  {journal}
  {\bibinfo  {journal} {Nature communications}\ }\textbf {\bibinfo {volume}
  {5}},\ \bibinfo {pages} {3786} (\bibinfo {year} {2014})}\BibitemShut
  {NoStop}%
\bibitem [{\citenamefont {Wan}\ \emph {et~al.}(2011)\citenamefont {Wan},
  \citenamefont {Turner}, \citenamefont {Vishwanath},\ and\ \citenamefont
  {Savrasov}}]{Wan2011}%
  \BibitemOpen
  \bibfield  {author} {\bibinfo {author} {\bibfnamefont {X.}~\bibnamefont
  {Wan}}, \bibinfo {author} {\bibfnamefont {A.~M.}\ \bibnamefont {Turner}},
  \bibinfo {author} {\bibfnamefont {A.}~\bibnamefont {Vishwanath}}, \ and\
  \bibinfo {author} {\bibfnamefont {S.~Y.}\ \bibnamefont {Savrasov}},\ }\href
  {\doibase 10.1103/PhysRevB.83.205101} {\bibfield  {journal} {\bibinfo
  {journal} {Phys. Rev. B}\ }\textbf {\bibinfo {volume} {83}},\ \bibinfo
  {pages} {205101} (\bibinfo {year} {2011})}\BibitemShut {NoStop}%
\bibitem [{\citenamefont {Burkov}\ \emph {et~al.}(2011)\citenamefont {Burkov},
  \citenamefont {Hook},\ and\ \citenamefont {Balents}}]{Burkov2011}%
  \BibitemOpen
  \bibfield  {author} {\bibinfo {author} {\bibfnamefont {A.~A.}\ \bibnamefont
  {Burkov}}, \bibinfo {author} {\bibfnamefont {M.~D.}\ \bibnamefont {Hook}}, \
  and\ \bibinfo {author} {\bibfnamefont {L.}~\bibnamefont {Balents}},\ }\href
  {\doibase 10.1103/PhysRevB.84.235126} {\bibfield  {journal} {\bibinfo
  {journal} {Phys. Rev. B}\ }\textbf {\bibinfo {volume} {84}},\ \bibinfo
  {pages} {235126} (\bibinfo {year} {2011})}\BibitemShut {NoStop}%
\bibitem [{\citenamefont {Vazifeh}\ and\ \citenamefont
  {Franz}(2013)}]{Vazifeh2013}%
  \BibitemOpen
  \bibfield  {author} {\bibinfo {author} {\bibfnamefont {M.~M.}\ \bibnamefont
  {Vazifeh}}\ and\ \bibinfo {author} {\bibfnamefont {M.}~\bibnamefont
  {Franz}},\ }\href {\doibase 10.1103/PhysRevLett.111.027201} {\bibfield
  {journal} {\bibinfo  {journal} {Phys. Rev. Lett.}\ }\textbf {\bibinfo
  {volume} {111}},\ \bibinfo {pages} {027201} (\bibinfo {year}
  {2013})}\BibitemShut {NoStop}%
\bibitem [{\citenamefont {Weng}\ \emph {et~al.}(2015)\citenamefont {Weng},
  \citenamefont {Fang}, \citenamefont {Fang}, \citenamefont {Bernevig},\ and\
  \citenamefont {Dai}}]{Weng2015}%
  \BibitemOpen
  \bibfield  {author} {\bibinfo {author} {\bibfnamefont {H.}~\bibnamefont
  {Weng}}, \bibinfo {author} {\bibfnamefont {C.}~\bibnamefont {Fang}}, \bibinfo
  {author} {\bibfnamefont {Z.}~\bibnamefont {Fang}}, \bibinfo {author}
  {\bibfnamefont {B.~A.}\ \bibnamefont {Bernevig}}, \ and\ \bibinfo {author}
  {\bibfnamefont {X.}~\bibnamefont {Dai}},\ }\href {\doibase
  10.1103/PhysRevX.5.011029} {\bibfield  {journal} {\bibinfo  {journal} {Phys.
  Rev. X}\ }\textbf {\bibinfo {volume} {5}},\ \bibinfo {pages} {011029}
  (\bibinfo {year} {2015})}\BibitemShut {NoStop}%
\bibitem [{\citenamefont {Huang}\ \emph
  {et~al.}(2015{\natexlab{a}})\citenamefont {Huang}, \citenamefont {Xu},
  \citenamefont {Belopolski}, \citenamefont {Lee}, \citenamefont {Chang},
  \citenamefont {Wang}, \citenamefont {Alidoust}, \citenamefont {Bian},
  \citenamefont {Neupane}, \citenamefont {Zhang}, \citenamefont {Jia},
  \citenamefont {Bansil}, \citenamefont {Lin},\ and\ \citenamefont
  {Hasan}}]{Huang2015}%
  \BibitemOpen
  \bibfield  {author} {\bibinfo {author} {\bibfnamefont {S.-M.}\ \bibnamefont
  {Huang}}, \bibinfo {author} {\bibfnamefont {S.-Y.}\ \bibnamefont {Xu}},
  \bibinfo {author} {\bibfnamefont {I.}~\bibnamefont {Belopolski}}, \bibinfo
  {author} {\bibfnamefont {C.-C.}\ \bibnamefont {Lee}}, \bibinfo {author}
  {\bibfnamefont {G.}~\bibnamefont {Chang}}, \bibinfo {author} {\bibfnamefont
  {B.}~\bibnamefont {Wang}}, \bibinfo {author} {\bibfnamefont {N.}~\bibnamefont
  {Alidoust}}, \bibinfo {author} {\bibfnamefont {G.}~\bibnamefont {Bian}},
  \bibinfo {author} {\bibfnamefont {M.}~\bibnamefont {Neupane}}, \bibinfo
  {author} {\bibfnamefont {C.}~\bibnamefont {Zhang}}, \bibinfo {author}
  {\bibfnamefont {S.}~\bibnamefont {Jia}}, \bibinfo {author} {\bibfnamefont
  {A.}~\bibnamefont {Bansil}}, \bibinfo {author} {\bibfnamefont
  {H.}~\bibnamefont {Lin}}, \ and\ \bibinfo {author} {\bibfnamefont
  {M.}~\bibnamefont {Hasan}},\ }\href
  {http://www.scopus.com/inward/record.url?eid=2-s2.0-84935480287&partnerID=40&md5=b1264d28de2c5c4429fd32bf3f9fefbb}
  {\bibfield  {journal} {\bibinfo  {journal} {Nature Communications}\ }\textbf
  {\bibinfo {volume} {6}},\ \bibinfo {pages} {7373} (\bibinfo {year}
  {2015}{\natexlab{a}})}\BibitemShut {NoStop}%
\bibitem [{\citenamefont {Xu}\ \emph {et~al.}(2015{\natexlab{a}})\citenamefont
  {Xu}, \citenamefont {Belopolski}, \citenamefont {Alidoust}, \citenamefont
  {Neupane}, \citenamefont {Bian}, \citenamefont {Zhang}, \citenamefont
  {Sankar}, \citenamefont {Chang}, \citenamefont {Yuan}, \citenamefont {Lee},
  \citenamefont {Huang}, \citenamefont {Zheng}, \citenamefont {Ma},
  \citenamefont {Sanchez}, \citenamefont {Wang}, \citenamefont {Bansil},
  \citenamefont {Chou}, \citenamefont {Shibayev}, \citenamefont {Lin},
  \citenamefont {Jia},\ and\ \citenamefont {Hasan}}]{Xu2015}%
  \BibitemOpen
  \bibfield  {author} {\bibinfo {author} {\bibfnamefont {S.-Y.}\ \bibnamefont
  {Xu}}, \bibinfo {author} {\bibfnamefont {I.}~\bibnamefont {Belopolski}},
  \bibinfo {author} {\bibfnamefont {N.}~\bibnamefont {Alidoust}}, \bibinfo
  {author} {\bibfnamefont {M.}~\bibnamefont {Neupane}}, \bibinfo {author}
  {\bibfnamefont {G.}~\bibnamefont {Bian}}, \bibinfo {author} {\bibfnamefont
  {C.}~\bibnamefont {Zhang}}, \bibinfo {author} {\bibfnamefont
  {R.}~\bibnamefont {Sankar}}, \bibinfo {author} {\bibfnamefont
  {G.}~\bibnamefont {Chang}}, \bibinfo {author} {\bibfnamefont
  {Z.}~\bibnamefont {Yuan}}, \bibinfo {author} {\bibfnamefont {C.-C.}\
  \bibnamefont {Lee}}, \bibinfo {author} {\bibfnamefont {S.-M.}\ \bibnamefont
  {Huang}}, \bibinfo {author} {\bibfnamefont {H.}~\bibnamefont {Zheng}},
  \bibinfo {author} {\bibfnamefont {J.}~\bibnamefont {Ma}}, \bibinfo {author}
  {\bibfnamefont {D.}~\bibnamefont {Sanchez}}, \bibinfo {author} {\bibfnamefont
  {B.}~\bibnamefont {Wang}}, \bibinfo {author} {\bibfnamefont {A.}~\bibnamefont
  {Bansil}}, \bibinfo {author} {\bibfnamefont {F.}~\bibnamefont {Chou}},
  \bibinfo {author} {\bibfnamefont {P.}~\bibnamefont {Shibayev}}, \bibinfo
  {author} {\bibfnamefont {H.}~\bibnamefont {Lin}}, \bibinfo {author}
  {\bibfnamefont {S.}~\bibnamefont {Jia}}, \ and\ \bibinfo {author}
  {\bibfnamefont {M.}~\bibnamefont {Hasan}},\ }\href {\doibase
  10.1126/science.aaa9297} {\bibfield  {journal} {\bibinfo  {journal}
  {Science}\ }\textbf {\bibinfo {volume} {349}},\ \bibinfo {pages} {613}
  (\bibinfo {year} {2015}{\natexlab{a}})}\BibitemShut {NoStop}%
\bibitem [{\citenamefont {Lv}\ \emph {et~al.}(2015{\natexlab{a}})\citenamefont
  {Lv}, \citenamefont {Weng}, \citenamefont {Fu}, \citenamefont {Wang},
  \citenamefont {Miao}, \citenamefont {Ma}, \citenamefont {Richard},
  \citenamefont {Huang}, \citenamefont {Zhao}, \citenamefont {Chen},
  \citenamefont {Fang}, \citenamefont {Dai}, \citenamefont {Qian},\ and\
  \citenamefont {Ding}}]{Lv2015}%
  \BibitemOpen
  \bibfield  {author} {\bibinfo {author} {\bibfnamefont {B.~Q.}\ \bibnamefont
  {Lv}}, \bibinfo {author} {\bibfnamefont {H.~M.}\ \bibnamefont {Weng}},
  \bibinfo {author} {\bibfnamefont {B.~B.}\ \bibnamefont {Fu}}, \bibinfo
  {author} {\bibfnamefont {X.~P.}\ \bibnamefont {Wang}}, \bibinfo {author}
  {\bibfnamefont {H.}~\bibnamefont {Miao}}, \bibinfo {author} {\bibfnamefont
  {J.}~\bibnamefont {Ma}}, \bibinfo {author} {\bibfnamefont {P.}~\bibnamefont
  {Richard}}, \bibinfo {author} {\bibfnamefont {X.~C.}\ \bibnamefont {Huang}},
  \bibinfo {author} {\bibfnamefont {L.~X.}\ \bibnamefont {Zhao}}, \bibinfo
  {author} {\bibfnamefont {G.~F.}\ \bibnamefont {Chen}}, \bibinfo {author}
  {\bibfnamefont {Z.}~\bibnamefont {Fang}}, \bibinfo {author} {\bibfnamefont
  {X.}~\bibnamefont {Dai}}, \bibinfo {author} {\bibfnamefont {T.}~\bibnamefont
  {Qian}}, \ and\ \bibinfo {author} {\bibfnamefont {H.}~\bibnamefont {Ding}},\
  }\href {\doibase 10.1103/PhysRevX.5.031013} {\bibfield  {journal} {\bibinfo
  {journal} {Phys. Rev. X}\ }\textbf {\bibinfo {volume} {5}},\ \bibinfo {pages}
  {031013} (\bibinfo {year} {2015}{\natexlab{a}})}\BibitemShut {NoStop}%
\bibitem [{\citenamefont {Xu}\ \emph {et~al.}(2015{\natexlab{b}})\citenamefont
  {Xu}, \citenamefont {Alidoust}, \citenamefont {Belopolski}, \citenamefont
  {Yuan}, \citenamefont {Bian}, \citenamefont {Chang}, \citenamefont {Zheng},
  \citenamefont {Strocov}, \citenamefont {Sanchez}, \citenamefont {Chang},
  \citenamefont {Zhang}, \citenamefont {Mou}, \citenamefont {Wu}, \citenamefont
  {Huang}, \citenamefont {Lee}, \citenamefont {Huang}, \citenamefont {Wang},
  \citenamefont {Bansil}, \citenamefont {Jeng}, \citenamefont {Neupert},
  \citenamefont {Kaminski}, \citenamefont {Lin}, \citenamefont {Jia},\ and\
  \citenamefont {Zahid~Hasan}}]{Xu20152}%
  \BibitemOpen
  \bibfield  {author} {\bibinfo {author} {\bibfnamefont {S.-Y.}\ \bibnamefont
  {Xu}}, \bibinfo {author} {\bibfnamefont {N.}~\bibnamefont {Alidoust}},
  \bibinfo {author} {\bibfnamefont {I.}~\bibnamefont {Belopolski}}, \bibinfo
  {author} {\bibfnamefont {Z.}~\bibnamefont {Yuan}}, \bibinfo {author}
  {\bibfnamefont {G.}~\bibnamefont {Bian}}, \bibinfo {author} {\bibfnamefont
  {T.-R.}\ \bibnamefont {Chang}}, \bibinfo {author} {\bibfnamefont
  {H.}~\bibnamefont {Zheng}}, \bibinfo {author} {\bibfnamefont
  {V.}~\bibnamefont {Strocov}}, \bibinfo {author} {\bibfnamefont
  {D.}~\bibnamefont {Sanchez}}, \bibinfo {author} {\bibfnamefont
  {G.}~\bibnamefont {Chang}}, \bibinfo {author} {\bibfnamefont
  {C.}~\bibnamefont {Zhang}}, \bibinfo {author} {\bibfnamefont
  {D.}~\bibnamefont {Mou}}, \bibinfo {author} {\bibfnamefont {Y.}~\bibnamefont
  {Wu}}, \bibinfo {author} {\bibfnamefont {L.}~\bibnamefont {Huang}}, \bibinfo
  {author} {\bibfnamefont {C.-C.}\ \bibnamefont {Lee}}, \bibinfo {author}
  {\bibfnamefont {S.-M.}\ \bibnamefont {Huang}}, \bibinfo {author}
  {\bibfnamefont {B.}~\bibnamefont {Wang}}, \bibinfo {author} {\bibfnamefont
  {A.}~\bibnamefont {Bansil}}, \bibinfo {author} {\bibfnamefont {H.-T.}\
  \bibnamefont {Jeng}}, \bibinfo {author} {\bibfnamefont {T.}~\bibnamefont
  {Neupert}}, \bibinfo {author} {\bibfnamefont {A.}~\bibnamefont {Kaminski}},
  \bibinfo {author} {\bibfnamefont {H.}~\bibnamefont {Lin}}, \bibinfo {author}
  {\bibfnamefont {S.}~\bibnamefont {Jia}}, \ and\ \bibinfo {author}
  {\bibfnamefont {M.}~\bibnamefont {Zahid~Hasan}},\ }\href
  {http://www.nature.com/nphys/journal/v11/n9/full/nphys3437.html} {\bibfield
  {journal} {\bibinfo  {journal} {Nature Physics}\ }\textbf {\bibinfo {volume}
  {11}},\ \bibinfo {pages} {748} (\bibinfo {year}
  {2015}{\natexlab{b}})}\BibitemShut {NoStop}%
\bibitem [{\citenamefont {Zhang}\ \emph {et~al.}(2017)\citenamefont {Zhang},
  \citenamefont {Yuan}, \citenamefont {Jiang}, \citenamefont {Tong},
  \citenamefont {Zhang}, \citenamefont {Xie},\ and\ \citenamefont
  {Jia}}]{Zhang2015}%
  \BibitemOpen
  \bibfield  {author} {\bibinfo {author} {\bibfnamefont {C.-L.}\ \bibnamefont
  {Zhang}}, \bibinfo {author} {\bibfnamefont {Z.}~\bibnamefont {Yuan}},
  \bibinfo {author} {\bibfnamefont {Q.-D.}\ \bibnamefont {Jiang}}, \bibinfo
  {author} {\bibfnamefont {B.}~\bibnamefont {Tong}}, \bibinfo {author}
  {\bibfnamefont {C.}~\bibnamefont {Zhang}}, \bibinfo {author} {\bibfnamefont
  {X.~C.}\ \bibnamefont {Xie}}, \ and\ \bibinfo {author} {\bibfnamefont
  {S.}~\bibnamefont {Jia}},\ }\href {\doibase 10.1103/PhysRevB.95.085202}
  {\bibfield  {journal} {\bibinfo  {journal} {Phys. Rev. B}\ }\textbf {\bibinfo
  {volume} {95}},\ \bibinfo {pages} {085202} (\bibinfo {year}
  {2017})}\BibitemShut {NoStop}%
\bibitem [{\citenamefont {Yang}\ \emph {et~al.}(2015)\citenamefont {Yang},
  \citenamefont {Liu}, \citenamefont {Sun}, \citenamefont {Peng}, \citenamefont
  {Yang}, \citenamefont {Zhang}, \citenamefont {Zhou}, \citenamefont {Zhang},
  \citenamefont {Guo}, \citenamefont {Rahn}, \citenamefont {Prabhakaran},
  \citenamefont {Hussain}, \citenamefont {Mo}, \citenamefont {Felser},
  \citenamefont {Yan},\ and\ \citenamefont {Chen}}]{Yang2015}%
  \BibitemOpen
  \bibfield  {author} {\bibinfo {author} {\bibfnamefont {L.}~\bibnamefont
  {Yang}}, \bibinfo {author} {\bibfnamefont {Z.}~\bibnamefont {Liu}}, \bibinfo
  {author} {\bibfnamefont {Y.}~\bibnamefont {Sun}}, \bibinfo {author}
  {\bibfnamefont {H.}~\bibnamefont {Peng}}, \bibinfo {author} {\bibfnamefont
  {H.}~\bibnamefont {Yang}}, \bibinfo {author} {\bibfnamefont {T.}~\bibnamefont
  {Zhang}}, \bibinfo {author} {\bibfnamefont {B.}~\bibnamefont {Zhou}},
  \bibinfo {author} {\bibfnamefont {Y.}~\bibnamefont {Zhang}}, \bibinfo
  {author} {\bibfnamefont {Y.}~\bibnamefont {Guo}}, \bibinfo {author}
  {\bibfnamefont {M.}~\bibnamefont {Rahn}}, \bibinfo {author} {\bibfnamefont
  {D.}~\bibnamefont {Prabhakaran}}, \bibinfo {author} {\bibfnamefont
  {Z.}~\bibnamefont {Hussain}}, \bibinfo {author} {\bibfnamefont {S.-K.}\
  \bibnamefont {Mo}}, \bibinfo {author} {\bibfnamefont {C.}~\bibnamefont
  {Felser}}, \bibinfo {author} {\bibfnamefont {B.}~\bibnamefont {Yan}}, \ and\
  \bibinfo {author} {\bibfnamefont {Y.}~\bibnamefont {Chen}},\ }\href
  {http://www.nature.com/nphys/journal/v11/n9/full/nphys3425.html} {\bibfield
  {journal} {\bibinfo  {journal} {Nature Physics}\ }\textbf {\bibinfo {volume}
  {11}},\ \bibinfo {pages} {728} (\bibinfo {year} {2015})}\BibitemShut
  {NoStop}%
\bibitem [{\citenamefont {Wang}\ \emph
  {et~al.}(2016{\natexlab{a}})\citenamefont {Wang}, \citenamefont {Zheng},
  \citenamefont {Shen}, \citenamefont {Lu}, \citenamefont {Fang}, \citenamefont
  {Sheng}, \citenamefont {Zhou}, \citenamefont {Yang}, \citenamefont {Li},
  \citenamefont {Feng},\ and\ \citenamefont {Xu}}]{Wang2015}%
  \BibitemOpen
  \bibfield  {author} {\bibinfo {author} {\bibfnamefont {Z.}~\bibnamefont
  {Wang}}, \bibinfo {author} {\bibfnamefont {Y.}~\bibnamefont {Zheng}},
  \bibinfo {author} {\bibfnamefont {Z.}~\bibnamefont {Shen}}, \bibinfo {author}
  {\bibfnamefont {Y.}~\bibnamefont {Lu}}, \bibinfo {author} {\bibfnamefont
  {H.}~\bibnamefont {Fang}}, \bibinfo {author} {\bibfnamefont {F.}~\bibnamefont
  {Sheng}}, \bibinfo {author} {\bibfnamefont {Y.}~\bibnamefont {Zhou}},
  \bibinfo {author} {\bibfnamefont {X.}~\bibnamefont {Yang}}, \bibinfo {author}
  {\bibfnamefont {Y.}~\bibnamefont {Li}}, \bibinfo {author} {\bibfnamefont
  {C.}~\bibnamefont {Feng}}, \ and\ \bibinfo {author} {\bibfnamefont {Z.-A.}\
  \bibnamefont {Xu}},\ }\href {\doibase 10.1103/PhysRevB.93.121112} {\bibfield
  {journal} {\bibinfo  {journal} {Phys. Rev. B}\ }\textbf {\bibinfo {volume}
  {93}},\ \bibinfo {pages} {121112} (\bibinfo {year}
  {2016}{\natexlab{a}})}\BibitemShut {NoStop}%
\bibitem [{\citenamefont {Huang}\ \emph
  {et~al.}(2015{\natexlab{b}})\citenamefont {Huang}, \citenamefont {Zhao},
  \citenamefont {Long}, \citenamefont {Wang}, \citenamefont {Chen},
  \citenamefont {Yang}, \citenamefont {Liang}, \citenamefont {Xue},
  \citenamefont {Weng}, \citenamefont {Fang}, \citenamefont {Dai},\ and\
  \citenamefont {Chen}}]{HuangXC2015}%
  \BibitemOpen
  \bibfield  {author} {\bibinfo {author} {\bibfnamefont {X.}~\bibnamefont
  {Huang}}, \bibinfo {author} {\bibfnamefont {L.}~\bibnamefont {Zhao}},
  \bibinfo {author} {\bibfnamefont {Y.}~\bibnamefont {Long}}, \bibinfo {author}
  {\bibfnamefont {P.}~\bibnamefont {Wang}}, \bibinfo {author} {\bibfnamefont
  {D.}~\bibnamefont {Chen}}, \bibinfo {author} {\bibfnamefont {Z.}~\bibnamefont
  {Yang}}, \bibinfo {author} {\bibfnamefont {H.}~\bibnamefont {Liang}},
  \bibinfo {author} {\bibfnamefont {M.}~\bibnamefont {Xue}}, \bibinfo {author}
  {\bibfnamefont {H.}~\bibnamefont {Weng}}, \bibinfo {author} {\bibfnamefont
  {Z.}~\bibnamefont {Fang}}, \bibinfo {author} {\bibfnamefont {X.}~\bibnamefont
  {Dai}}, \ and\ \bibinfo {author} {\bibfnamefont {G.}~\bibnamefont {Chen}},\
  }\href {\doibase 10.1103/PhysRevX.5.031023} {\bibfield  {journal} {\bibinfo
  {journal} {Phys. Rev. X}\ }\textbf {\bibinfo {volume} {5}},\ \bibinfo {pages}
  {031023} (\bibinfo {year} {2015}{\natexlab{b}})}\BibitemShut {NoStop}%
\bibitem [{\citenamefont {Lv}\ \emph {et~al.}(2015{\natexlab{b}})\citenamefont
  {Lv}, \citenamefont {Xu}, \citenamefont {Weng}, \citenamefont {Ma},
  \citenamefont {Richard}, \citenamefont {Huang}, \citenamefont {Zhao},
  \citenamefont {Chen}, \citenamefont {Matt}, \citenamefont {Bisti},
  \citenamefont {Strocov}, \citenamefont {Mesot}, \citenamefont {Fang},
  \citenamefont {Dai}, \citenamefont {Qian}, \citenamefont {Shi},\ and\
  \citenamefont {Ding}}]{Lv2015724}%
  \BibitemOpen
  \bibfield  {author} {\bibinfo {author} {\bibfnamefont {B.}~\bibnamefont
  {Lv}}, \bibinfo {author} {\bibfnamefont {N.}~\bibnamefont {Xu}}, \bibinfo
  {author} {\bibfnamefont {H.}~\bibnamefont {Weng}}, \bibinfo {author}
  {\bibfnamefont {J.}~\bibnamefont {Ma}}, \bibinfo {author} {\bibfnamefont
  {P.}~\bibnamefont {Richard}}, \bibinfo {author} {\bibfnamefont
  {X.}~\bibnamefont {Huang}}, \bibinfo {author} {\bibfnamefont
  {L.}~\bibnamefont {Zhao}}, \bibinfo {author} {\bibfnamefont {G.}~\bibnamefont
  {Chen}}, \bibinfo {author} {\bibfnamefont {C.}~\bibnamefont {Matt}}, \bibinfo
  {author} {\bibfnamefont {F.}~\bibnamefont {Bisti}}, \bibinfo {author}
  {\bibfnamefont {V.}~\bibnamefont {Strocov}}, \bibinfo {author} {\bibfnamefont
  {J.}~\bibnamefont {Mesot}}, \bibinfo {author} {\bibfnamefont
  {Z.}~\bibnamefont {Fang}}, \bibinfo {author} {\bibfnamefont {X.}~\bibnamefont
  {Dai}}, \bibinfo {author} {\bibfnamefont {T.}~\bibnamefont {Qian}}, \bibinfo
  {author} {\bibfnamefont {M.}~\bibnamefont {Shi}}, \ and\ \bibinfo {author}
  {\bibfnamefont {H.}~\bibnamefont {Ding}},\ }\href {\doibase
  10.1038/nphys3426} {\bibfield  {journal} {\bibinfo  {journal} {Nature
  Physics}\ }\textbf {\bibinfo {volume} {11}},\ \bibinfo {pages} {724}
  (\bibinfo {year} {2015}{\natexlab{b}})}\BibitemShut {NoStop}%
\bibitem [{\citenamefont {Lopez-Bezanilla}\ and\ \citenamefont
  {Littlewood}(2016)}]{Lopez2016}%
  \BibitemOpen
  \bibfield  {author} {\bibinfo {author} {\bibfnamefont {A.}~\bibnamefont
  {Lopez-Bezanilla}}\ and\ \bibinfo {author} {\bibfnamefont {P.~B.}\
  \bibnamefont {Littlewood}},\ }\href {\doibase 10.1103/PhysRevB.93.241405}
  {\bibfield  {journal} {\bibinfo  {journal} {Phys. Rev. B}\ }\textbf {\bibinfo
  {volume} {93}},\ \bibinfo {pages} {241405} (\bibinfo {year}
  {2016})}\BibitemShut {NoStop}%
\bibitem [{\citenamefont {Goerbig}\ \emph {et~al.}(2008)\citenamefont
  {Goerbig}, \citenamefont {Fuchs}, \citenamefont {Montambaux},\ and\
  \citenamefont {Pi\'echon}}]{Goerbig2008}%
  \BibitemOpen
  \bibfield  {author} {\bibinfo {author} {\bibfnamefont {M.~O.}\ \bibnamefont
  {Goerbig}}, \bibinfo {author} {\bibfnamefont {J.-N.}\ \bibnamefont {Fuchs}},
  \bibinfo {author} {\bibfnamefont {G.}~\bibnamefont {Montambaux}}, \ and\
  \bibinfo {author} {\bibfnamefont {F.}~\bibnamefont {Pi\'echon}},\ }\href
  {\doibase 10.1103/PhysRevB.78.045415} {\bibfield  {journal} {\bibinfo
  {journal} {Phys. Rev. B}\ }\textbf {\bibinfo {volume} {78}},\ \bibinfo
  {pages} {045415} (\bibinfo {year} {2008})}\BibitemShut {NoStop}%
\bibitem [{\citenamefont {Hirata}\ \emph {et~al.}(2017)\citenamefont {Hirata},
  \citenamefont {Ishikawa}, \citenamefont {Matsuno}, \citenamefont {Kobayashi},
  \citenamefont {Miyagawa}, \citenamefont {Tamura}, \citenamefont {Berthier},\
  and\ \citenamefont {Kanoda}}]{Hirata2017}%
  \BibitemOpen
  \bibfield  {author} {\bibinfo {author} {\bibfnamefont {M.}~\bibnamefont
  {Hirata}}, \bibinfo {author} {\bibfnamefont {K.}~\bibnamefont {Ishikawa}},
  \bibinfo {author} {\bibfnamefont {G.}~\bibnamefont {Matsuno}}, \bibinfo
  {author} {\bibfnamefont {A.}~\bibnamefont {Kobayashi}}, \bibinfo {author}
  {\bibfnamefont {K.}~\bibnamefont {Miyagawa}}, \bibinfo {author}
  {\bibfnamefont {M.}~\bibnamefont {Tamura}}, \bibinfo {author} {\bibfnamefont
  {C.}~\bibnamefont {Berthier}}, \ and\ \bibinfo {author} {\bibfnamefont
  {K.}~\bibnamefont {Kanoda}},\ }\href {\doibase 10.1126/science.aan5351}
  {\bibfield  {journal} {\bibinfo  {journal} {Science}\ }\textbf {\bibinfo
  {volume} {358}},\ \bibinfo {pages} {1403} (\bibinfo {year}
  {2017})}\BibitemShut {NoStop}%
\bibitem [{\citenamefont {Soluyanov}(2017)}]{soluyanov2017}%
  \BibitemOpen
  \bibfield  {author} {\bibinfo {author} {\bibfnamefont {A.~A.}\ \bibnamefont
  {Soluyanov}},\ }\href@noop {} {\bibfield  {journal} {\bibinfo  {journal}
  {Physics}\ }\textbf {\bibinfo {volume} {10}},\ \bibinfo {pages} {74}
  (\bibinfo {year} {2017})}\BibitemShut {NoStop}%
\bibitem [{\citenamefont {Soluyanov}\ \emph {et~al.}(2015)\citenamefont
  {Soluyanov}, \citenamefont {Gresch}, \citenamefont {Wang}, \citenamefont
  {Wu}, \citenamefont {Troyer}, \citenamefont {Dai},\ and\ \citenamefont
  {Bernevig}}]{soluyanov2015}%
  \BibitemOpen
  \bibfield  {author} {\bibinfo {author} {\bibfnamefont {A.~A.}\ \bibnamefont
  {Soluyanov}}, \bibinfo {author} {\bibfnamefont {D.}~\bibnamefont {Gresch}},
  \bibinfo {author} {\bibfnamefont {Z.}~\bibnamefont {Wang}}, \bibinfo {author}
  {\bibfnamefont {Q.}~\bibnamefont {Wu}}, \bibinfo {author} {\bibfnamefont
  {M.}~\bibnamefont {Troyer}}, \bibinfo {author} {\bibfnamefont
  {X.}~\bibnamefont {Dai}}, \ and\ \bibinfo {author} {\bibfnamefont {B.~A.}\
  \bibnamefont {Bernevig}},\ }\href
  {https://www.nature.com/nature/journal/v527/n7579/full/nature15768.html#ref23}
  {\bibfield  {journal} {\bibinfo  {journal} {Nature (London)}\ }\textbf
  {\bibinfo {volume} {527}},\ \bibinfo {pages} {495} (\bibinfo {year}
  {2015})}\BibitemShut {NoStop}%
\bibitem [{\citenamefont {Xu}\ \emph {et~al.}(2015{\natexlab{c}})\citenamefont
  {Xu}, \citenamefont {Zhang},\ and\ \citenamefont {Zhang}}]{Xu2015prl}%
  \BibitemOpen
  \bibfield  {author} {\bibinfo {author} {\bibfnamefont {Y.}~\bibnamefont
  {Xu}}, \bibinfo {author} {\bibfnamefont {F.}~\bibnamefont {Zhang}}, \ and\
  \bibinfo {author} {\bibfnamefont {C.}~\bibnamefont {Zhang}},\ }\href
  {\doibase 10.1103/PhysRevLett.115.265304} {\bibfield  {journal} {\bibinfo
  {journal} {Phys. Rev. Lett.}\ }\textbf {\bibinfo {volume} {115}},\ \bibinfo
  {pages} {265304} (\bibinfo {year} {2015}{\natexlab{c}})}\BibitemShut
  {NoStop}%
\bibitem [{\citenamefont {Sun}\ \emph {et~al.}(2015{\natexlab{a}})\citenamefont
  {Sun}, \citenamefont {Wu}, \citenamefont {Ali}, \citenamefont {Felser},\ and\
  \citenamefont {Yan}}]{Sunyan2015}%
  \BibitemOpen
  \bibfield  {author} {\bibinfo {author} {\bibfnamefont {Y.}~\bibnamefont
  {Sun}}, \bibinfo {author} {\bibfnamefont {S.-C.}\ \bibnamefont {Wu}},
  \bibinfo {author} {\bibfnamefont {M.~N.}\ \bibnamefont {Ali}}, \bibinfo
  {author} {\bibfnamefont {C.}~\bibnamefont {Felser}}, \ and\ \bibinfo {author}
  {\bibfnamefont {B.}~\bibnamefont {Yan}},\ }\href {\doibase
  10.1103/PhysRevB.92.161107} {\bibfield  {journal} {\bibinfo  {journal} {Phys.
  Rev. B}\ }\textbf {\bibinfo {volume} {92}},\ \bibinfo {pages} {161107}
  (\bibinfo {year} {2015}{\natexlab{a}})}\BibitemShut {NoStop}%
\bibitem [{\citenamefont {Wang}\ \emph
  {et~al.}(2016{\natexlab{b}})\citenamefont {Wang}, \citenamefont {Gresch},
  \citenamefont {Soluyanov}, \citenamefont {Xie}, \citenamefont {Kushwaha},
  \citenamefont {Dai}, \citenamefont {Troyer}, \citenamefont {Cava},\ and\
  \citenamefont {Bernevig}}]{Wang2016}%
  \BibitemOpen
  \bibfield  {author} {\bibinfo {author} {\bibfnamefont {Z.}~\bibnamefont
  {Wang}}, \bibinfo {author} {\bibfnamefont {D.}~\bibnamefont {Gresch}},
  \bibinfo {author} {\bibfnamefont {A.~A.}\ \bibnamefont {Soluyanov}}, \bibinfo
  {author} {\bibfnamefont {W.}~\bibnamefont {Xie}}, \bibinfo {author}
  {\bibfnamefont {S.}~\bibnamefont {Kushwaha}}, \bibinfo {author}
  {\bibfnamefont {X.}~\bibnamefont {Dai}}, \bibinfo {author} {\bibfnamefont
  {M.}~\bibnamefont {Troyer}}, \bibinfo {author} {\bibfnamefont {R.~J.}\
  \bibnamefont {Cava}}, \ and\ \bibinfo {author} {\bibfnamefont {B.~A.}\
  \bibnamefont {Bernevig}},\ }\href {\doibase 10.1103/PhysRevLett.117.056805}
  {\bibfield  {journal} {\bibinfo  {journal} {Phys. Rev. Lett.}\ }\textbf
  {\bibinfo {volume} {117}},\ \bibinfo {pages} {056805} (\bibinfo {year}
  {2016}{\natexlab{b}})}\BibitemShut {NoStop}%
\bibitem [{\citenamefont {Deng}\ \emph {et~al.}(2016)\citenamefont {Deng},
  \citenamefont {Wan}, \citenamefont {Deng}, \citenamefont {Zhang},
  \citenamefont {Ding}, \citenamefont {Wang}, \citenamefont {Yan},
  \citenamefont {Huang}, \citenamefont {Zhang}, \citenamefont {Xu} \emph
  {et~al.}}]{Deng2016}%
  \BibitemOpen
  \bibfield  {author} {\bibinfo {author} {\bibfnamefont {K.}~\bibnamefont
  {Deng}}, \bibinfo {author} {\bibfnamefont {G.}~\bibnamefont {Wan}}, \bibinfo
  {author} {\bibfnamefont {P.}~\bibnamefont {Deng}}, \bibinfo {author}
  {\bibfnamefont {K.}~\bibnamefont {Zhang}}, \bibinfo {author} {\bibfnamefont
  {S.}~\bibnamefont {Ding}}, \bibinfo {author} {\bibfnamefont {E.}~\bibnamefont
  {Wang}}, \bibinfo {author} {\bibfnamefont {M.}~\bibnamefont {Yan}}, \bibinfo
  {author} {\bibfnamefont {H.}~\bibnamefont {Huang}}, \bibinfo {author}
  {\bibfnamefont {H.}~\bibnamefont {Zhang}}, \bibinfo {author} {\bibfnamefont
  {Z.}~\bibnamefont {Xu}},  \emph {et~al.},\ }\href
  {https://www.nature.com/articles/nphys3871} {\bibfield  {journal} {\bibinfo
  {journal} {Nature Physics}\ }\textbf {\bibinfo {volume} {12}},\ \bibinfo
  {pages} {1105} (\bibinfo {year} {2016})}\BibitemShut {NoStop}%
\bibitem [{\citenamefont {Huang}\ \emph {et~al.}(2016)\citenamefont {Huang},
  \citenamefont {McCormick}, \citenamefont {Ochi}, \citenamefont {Zhao},
  \citenamefont {Suzuki}, \citenamefont {Arita}, \citenamefont {Wu},
  \citenamefont {Mou}, \citenamefont {Cao}, \citenamefont {Yan} \emph
  {et~al.}}]{Huang2016}%
  \BibitemOpen
  \bibfield  {author} {\bibinfo {author} {\bibfnamefont {L.}~\bibnamefont
  {Huang}}, \bibinfo {author} {\bibfnamefont {T.~M.}\ \bibnamefont
  {McCormick}}, \bibinfo {author} {\bibfnamefont {M.}~\bibnamefont {Ochi}},
  \bibinfo {author} {\bibfnamefont {Z.}~\bibnamefont {Zhao}}, \bibinfo {author}
  {\bibfnamefont {M.-T.}\ \bibnamefont {Suzuki}}, \bibinfo {author}
  {\bibfnamefont {R.}~\bibnamefont {Arita}}, \bibinfo {author} {\bibfnamefont
  {Y.}~\bibnamefont {Wu}}, \bibinfo {author} {\bibfnamefont {D.}~\bibnamefont
  {Mou}}, \bibinfo {author} {\bibfnamefont {H.}~\bibnamefont {Cao}}, \bibinfo
  {author} {\bibfnamefont {J.}~\bibnamefont {Yan}},  \emph {et~al.},\ }\href
  {https://www.nature.com/articles/nmat4685} {\bibfield  {journal} {\bibinfo
  {journal} {Nature materials}\ }\textbf {\bibinfo {volume} {15}},\ \bibinfo
  {pages} {1155} (\bibinfo {year} {2016})}\BibitemShut {NoStop}%
\bibitem [{\citenamefont {Jiang}\ \emph {et~al.}(2017)\citenamefont {Jiang},
  \citenamefont {Liu}, \citenamefont {Sun}, \citenamefont {Yang}, \citenamefont
  {Rajamathi}, \citenamefont {Qi}, \citenamefont {Yang}, \citenamefont {Chen},
  \citenamefont {Peng}, \citenamefont {Hwang} \emph {et~al.}}]{Jiang2017}%
  \BibitemOpen
  \bibfield  {author} {\bibinfo {author} {\bibfnamefont {J.}~\bibnamefont
  {Jiang}}, \bibinfo {author} {\bibfnamefont {Z.}~\bibnamefont {Liu}}, \bibinfo
  {author} {\bibfnamefont {Y.}~\bibnamefont {Sun}}, \bibinfo {author}
  {\bibfnamefont {H.}~\bibnamefont {Yang}}, \bibinfo {author} {\bibfnamefont
  {C.}~\bibnamefont {Rajamathi}}, \bibinfo {author} {\bibfnamefont
  {Y.}~\bibnamefont {Qi}}, \bibinfo {author} {\bibfnamefont {L.}~\bibnamefont
  {Yang}}, \bibinfo {author} {\bibfnamefont {C.}~\bibnamefont {Chen}}, \bibinfo
  {author} {\bibfnamefont {H.}~\bibnamefont {Peng}}, \bibinfo {author}
  {\bibfnamefont {C.}~\bibnamefont {Hwang}},  \emph {et~al.},\ }\href
  {https://www.nature.com/articles/ncomms13973} {\bibfield  {journal} {\bibinfo
   {journal} {Nature communications}\ }\textbf {\bibinfo {volume} {8}},\
  \bibinfo {pages} {13973} (\bibinfo {year} {2017})}\BibitemShut {NoStop}%
\bibitem [{\citenamefont {Chang}\ \emph {et~al.}(2016)\citenamefont {Chang},
  \citenamefont {Xu}, \citenamefont {Sanchez}, \citenamefont {Huang},
  \citenamefont {Lee}, \citenamefont {Chang}, \citenamefont {Bian},
  \citenamefont {Zheng}, \citenamefont {Belopolski}, \citenamefont {Alidoust}
  \emph {et~al.}}]{chang2016}%
  \BibitemOpen
  \bibfield  {author} {\bibinfo {author} {\bibfnamefont {G.}~\bibnamefont
  {Chang}}, \bibinfo {author} {\bibfnamefont {S.-Y.}\ \bibnamefont {Xu}},
  \bibinfo {author} {\bibfnamefont {D.~S.}\ \bibnamefont {Sanchez}}, \bibinfo
  {author} {\bibfnamefont {S.-M.}\ \bibnamefont {Huang}}, \bibinfo {author}
  {\bibfnamefont {C.-C.}\ \bibnamefont {Lee}}, \bibinfo {author} {\bibfnamefont
  {T.-R.}\ \bibnamefont {Chang}}, \bibinfo {author} {\bibfnamefont
  {G.}~\bibnamefont {Bian}}, \bibinfo {author} {\bibfnamefont {H.}~\bibnamefont
  {Zheng}}, \bibinfo {author} {\bibfnamefont {I.}~\bibnamefont {Belopolski}},
  \bibinfo {author} {\bibfnamefont {N.}~\bibnamefont {Alidoust}},  \emph
  {et~al.},\ }\href {http://advances.sciencemag.org/content/2/6/e1600295}
  {\bibfield  {journal} {\bibinfo  {journal} {Science Advances}\ }\textbf
  {\bibinfo {volume} {2}},\ \bibinfo {pages} {e1600295} (\bibinfo {year}
  {2016})}\BibitemShut {NoStop}%
\bibitem [{\citenamefont {Xu~S-Y}(2017)}]{XuSY2017}%
  \BibitemOpen
  \bibfield  {author} {\bibinfo {author} {\bibfnamefont {C.~G. e.~a.}\
  \bibnamefont {Xu~S-Y}, \bibfnamefont {Alidoust~N}},\ }\href
  {https://www.ncbi.nlm.nih.gov/pmc/articles/PMC5457030/} {\bibfield  {journal}
  {\bibinfo  {journal} {Science Advances}\ }\textbf {\bibinfo {volume} {3
  (6)}},\ \bibinfo {pages} {e1603266} (\bibinfo {year} {2017})}\BibitemShut
  {NoStop}%
\bibitem [{\citenamefont {O'Brien}\ \emph {et~al.}(2016)\citenamefont
  {O'Brien}, \citenamefont {Diez},\ and\ \citenamefont
  {Beenakker}}]{O’Brien2016}%
  \BibitemOpen
  \bibfield  {author} {\bibinfo {author} {\bibfnamefont {T.~E.}\ \bibnamefont
  {O'Brien}}, \bibinfo {author} {\bibfnamefont {M.}~\bibnamefont {Diez}}, \
  and\ \bibinfo {author} {\bibfnamefont {C.~W.~J.}\ \bibnamefont {Beenakker}},\
  }\href {\doibase 10.1103/PhysRevLett.116.236401} {\bibfield  {journal}
  {\bibinfo  {journal} {Phys. Rev. Lett.}\ }\textbf {\bibinfo {volume} {116}},\
  \bibinfo {pages} {236401} (\bibinfo {year} {2016})}\BibitemShut {NoStop}%
\bibitem [{\citenamefont {Yu}\ \emph {et~al.}(2016)\citenamefont {Yu},
  \citenamefont {Yao},\ and\ \citenamefont {Yang}}]{Yang2016}%
  \BibitemOpen
  \bibfield  {author} {\bibinfo {author} {\bibfnamefont {Z.-M.}\ \bibnamefont
  {Yu}}, \bibinfo {author} {\bibfnamefont {Y.}~\bibnamefont {Yao}}, \ and\
  \bibinfo {author} {\bibfnamefont {S.~A.}\ \bibnamefont {Yang}},\ }\href
  {\doibase 10.1103/PhysRevLett.117.077202} {\bibfield  {journal} {\bibinfo
  {journal} {Phys. Rev. Lett.}\ }\textbf {\bibinfo {volume} {117}},\ \bibinfo
  {pages} {077202} (\bibinfo {year} {2016})}\BibitemShut {NoStop}%
\bibitem [{\citenamefont {Tchoumakov}\ \emph {et~al.}(2016)\citenamefont
  {Tchoumakov}, \citenamefont {Civelli},\ and\ \citenamefont
  {Goerbig}}]{Tchoumakov2016}%
  \BibitemOpen
  \bibfield  {author} {\bibinfo {author} {\bibfnamefont {S.}~\bibnamefont
  {Tchoumakov}}, \bibinfo {author} {\bibfnamefont {M.}~\bibnamefont {Civelli}},
  \ and\ \bibinfo {author} {\bibfnamefont {M.~O.}\ \bibnamefont {Goerbig}},\
  }\href {\doibase 10.1103/PhysRevLett.117.086402} {\bibfield  {journal}
  {\bibinfo  {journal} {Phys. Rev. Lett.}\ }\textbf {\bibinfo {volume} {117}},\
  \bibinfo {pages} {086402} (\bibinfo {year} {2016})}\BibitemShut {NoStop}%
\bibitem [{\citenamefont {Udagawa}\ and\ \citenamefont
  {Bergholtz}(2016)}]{Udagawa2016}%
  \BibitemOpen
  \bibfield  {author} {\bibinfo {author} {\bibfnamefont {M.}~\bibnamefont
  {Udagawa}}\ and\ \bibinfo {author} {\bibfnamefont {E.~J.}\ \bibnamefont
  {Bergholtz}},\ }\href {\doibase 10.1103/PhysRevLett.117.086401} {\bibfield
  {journal} {\bibinfo  {journal} {Phys. Rev. Lett.}\ }\textbf {\bibinfo
  {volume} {117}},\ \bibinfo {pages} {086401} (\bibinfo {year}
  {2016})}\BibitemShut {NoStop}%
\bibitem [{\citenamefont {Ferreiros}\ \emph {et~al.}(2017)\citenamefont
  {Ferreiros}, \citenamefont {Zyuzin},\ and\ \citenamefont
  {Bardarson}}]{Ferreiros2017}%
  \BibitemOpen
  \bibfield  {author} {\bibinfo {author} {\bibfnamefont {Y.}~\bibnamefont
  {Ferreiros}}, \bibinfo {author} {\bibfnamefont {A.~A.}\ \bibnamefont
  {Zyuzin}}, \ and\ \bibinfo {author} {\bibfnamefont {J.~H.}\ \bibnamefont
  {Bardarson}},\ }\href {\doibase 10.1103/PhysRevB.96.115202} {\bibfield
  {journal} {\bibinfo  {journal} {Phys. Rev. B}\ }\textbf {\bibinfo {volume}
  {96}},\ \bibinfo {pages} {115202} (\bibinfo {year} {2017})}\BibitemShut
  {NoStop}%
\bibitem [{\citenamefont {Saha}\ and\ \citenamefont {Tewari}(2018)}]{saha2018}%
  \BibitemOpen
  \bibfield  {author} {\bibinfo {author} {\bibfnamefont {S.}~\bibnamefont
  {Saha}}\ and\ \bibinfo {author} {\bibfnamefont {S.}~\bibnamefont {Tewari}},\
  }\href {https://link.springer.com/article/10.1140%2Fepjb%2Fe2017-80437-4}
  {\bibfield  {journal} {\bibinfo  {journal} {The European Physical Journal B}\
  }\textbf {\bibinfo {volume} {91}},\ \bibinfo {pages} {4} (\bibinfo {year}
  {2018})}\BibitemShut {NoStop}%
\bibitem [{\citenamefont {Krishna-murthy}\ \emph {et~al.}(1980)\citenamefont
  {Krishna-murthy}, \citenamefont {Wilkins},\ and\ \citenamefont
  {Wilson}}]{Krishna1980}%
  \BibitemOpen
  \bibfield  {author} {\bibinfo {author} {\bibfnamefont {H.~R.}\ \bibnamefont
  {Krishna-murthy}}, \bibinfo {author} {\bibfnamefont {J.~W.}\ \bibnamefont
  {Wilkins}}, \ and\ \bibinfo {author} {\bibfnamefont {K.~G.}\ \bibnamefont
  {Wilson}},\ }\href {\doibase 10.1103/PhysRevB.21.1003} {\bibfield  {journal}
  {\bibinfo  {journal} {Phys. Rev. B}\ }\textbf {\bibinfo {volume} {21}},\
  \bibinfo {pages} {1003} (\bibinfo {year} {1980})}\BibitemShut {NoStop}%
\bibitem [{\citenamefont {Tsvelick}\ and\ \citenamefont
  {Wiegmann}(1984)}]{tsvelick1984}%
  \BibitemOpen
  \bibfield  {author} {\bibinfo {author} {\bibfnamefont {A.}~\bibnamefont
  {Tsvelick}}\ and\ \bibinfo {author} {\bibfnamefont {P.}~\bibnamefont
  {Wiegmann}},\ }\href@noop {} {\bibfield  {journal} {\bibinfo  {journal}
  {Zeitschrift f{\"u}r Physik B Condensed Matter}\ }\textbf {\bibinfo {volume}
  {54}},\ \bibinfo {pages} {201} (\bibinfo {year} {1984})}\BibitemShut
  {NoStop}%
\bibitem [{\citenamefont {Andrei}\ and\ \citenamefont
  {Destri}(1984)}]{Andrei1984}%
  \BibitemOpen
  \bibfield  {author} {\bibinfo {author} {\bibfnamefont {N.}~\bibnamefont
  {Andrei}}\ and\ \bibinfo {author} {\bibfnamefont {C.}~\bibnamefont
  {Destri}},\ }\href {\doibase 10.1103/PhysRevLett.52.364} {\bibfield
  {journal} {\bibinfo  {journal} {Phys. Rev. Lett.}\ }\textbf {\bibinfo
  {volume} {52}},\ \bibinfo {pages} {364} (\bibinfo {year} {1984})}\BibitemShut
  {NoStop}%
\bibitem [{\citenamefont {Zhang}\ and\ \citenamefont {Lee}(1983)}]{Zhang1983}%
  \BibitemOpen
  \bibfield  {author} {\bibinfo {author} {\bibfnamefont {F.~C.}\ \bibnamefont
  {Zhang}}\ and\ \bibinfo {author} {\bibfnamefont {T.~K.}\ \bibnamefont
  {Lee}},\ }\href {\doibase 10.1103/PhysRevB.28.33} {\bibfield  {journal}
  {\bibinfo  {journal} {Phys. Rev. B}\ }\textbf {\bibinfo {volume} {28}},\
  \bibinfo {pages} {33} (\bibinfo {year} {1983})}\BibitemShut {NoStop}%
\bibitem [{\citenamefont {Coleman}(1984)}]{Coleman1984}%
  \BibitemOpen
  \bibfield  {author} {\bibinfo {author} {\bibfnamefont {P.}~\bibnamefont
  {Coleman}},\ }\href {\doibase 10.1103/PhysRevB.29.3035} {\bibfield  {journal}
  {\bibinfo  {journal} {Phys. Rev. B}\ }\textbf {\bibinfo {volume} {29}},\
  \bibinfo {pages} {3035} (\bibinfo {year} {1984})}\BibitemShut {NoStop}%
\bibitem [{\citenamefont {Read}\ and\ \citenamefont {Newns}(1983)}]{read1983}%
  \BibitemOpen
  \bibfield  {author} {\bibinfo {author} {\bibfnamefont {N.}~\bibnamefont
  {Read}}\ and\ \bibinfo {author} {\bibfnamefont {D.}~\bibnamefont {Newns}},\
  }\href@noop {} {\bibfield  {journal} {\bibinfo  {journal} {Journal of Physics
  C: Solid State Physics}\ }\textbf {\bibinfo {volume} {16}},\ \bibinfo {pages}
  {3273} (\bibinfo {year} {1983})}\BibitemShut {NoStop}%
\bibitem [{\citenamefont {Kuramoto}(1983)}]{Kuramoto1983}%
  \BibitemOpen
  \bibfield  {author} {\bibinfo {author} {\bibfnamefont {Y.}~\bibnamefont
  {Kuramoto}},\ }\href {\doibase 10.1007/BF01578246} {\bibfield  {journal}
  {\bibinfo  {journal} {Zeitschrift f$\ddot{u}$r Physik B Condensed Matter}\
  }\textbf {\bibinfo {volume} {53}},\ \bibinfo {pages} {37} (\bibinfo {year}
  {1983})}\BibitemShut {NoStop}%
\bibitem [{\citenamefont {Gunnarsson}\ and\ \citenamefont
  {Sch\"onhammer}(1983)}]{Gunnarsson1983}%
  \BibitemOpen
  \bibfield  {author} {\bibinfo {author} {\bibfnamefont {O.}~\bibnamefont
  {Gunnarsson}}\ and\ \bibinfo {author} {\bibfnamefont {K.}~\bibnamefont
  {Sch\"onhammer}},\ }\href {\doibase 10.1103/PhysRevLett.50.604} {\bibfield
  {journal} {\bibinfo  {journal} {Phys. Rev. Lett.}\ }\textbf {\bibinfo
  {volume} {50}},\ \bibinfo {pages} {604} (\bibinfo {year} {1983})}\BibitemShut
  {NoStop}%
\bibitem [{\citenamefont {Affleck}(1990)}]{affleck1990}%
  \BibitemOpen
  \bibfield  {author} {\bibinfo {author} {\bibfnamefont {I.}~\bibnamefont
  {Affleck}},\ }\href@noop {} {\bibfield  {journal} {\bibinfo  {journal}
  {Nuclear Physics B}\ }\textbf {\bibinfo {volume} {336}},\ \bibinfo {pages}
  {517} (\bibinfo {year} {1990})}\BibitemShut {NoStop}%
\bibitem [{\citenamefont {Gonzalez-Buxton}\ and\ \citenamefont
  {Ingersent}(1998)}]{Gonzalez1998}%
  \BibitemOpen
  \bibfield  {author} {\bibinfo {author} {\bibfnamefont {C.}~\bibnamefont
  {Gonzalez-Buxton}}\ and\ \bibinfo {author} {\bibfnamefont {K.}~\bibnamefont
  {Ingersent}},\ }\href {\doibase 10.1103/PhysRevB.57.14254} {\bibfield
  {journal} {\bibinfo  {journal} {Phys. Rev. B}\ }\textbf {\bibinfo {volume}
  {57}},\ \bibinfo {pages} {14254} (\bibinfo {year} {1998})}\BibitemShut
  {NoStop}%
\bibitem [{\citenamefont {Fritz}\ and\ \citenamefont
  {Vojta}(2004)}]{Fritz2004}%
  \BibitemOpen
  \bibfield  {author} {\bibinfo {author} {\bibfnamefont {L.}~\bibnamefont
  {Fritz}}\ and\ \bibinfo {author} {\bibfnamefont {M.}~\bibnamefont {Vojta}},\
  }\href {\doibase 10.1103/PhysRevB.70.214427} {\bibfield  {journal} {\bibinfo
  {journal} {Phys. Rev. B}\ }\textbf {\bibinfo {volume} {70}},\ \bibinfo
  {pages} {214427} (\bibinfo {year} {2004})}\BibitemShut {NoStop}%
\bibitem [{\citenamefont {Vojta}\ and\ \citenamefont
  {Fritz}(2004)}]{Vojta2004}%
  \BibitemOpen
  \bibfield  {author} {\bibinfo {author} {\bibfnamefont {M.}~\bibnamefont
  {Vojta}}\ and\ \bibinfo {author} {\bibfnamefont {L.}~\bibnamefont {Fritz}},\
  }\href {\doibase 10.1103/PhysRevB.70.094502} {\bibfield  {journal} {\bibinfo
  {journal} {Phys. Rev. B}\ }\textbf {\bibinfo {volume} {70}},\ \bibinfo
  {pages} {094502} (\bibinfo {year} {2004})}\BibitemShut {NoStop}%
\bibitem [{\citenamefont {Chang}\ \emph {et~al.}(2015)\citenamefont {Chang},
  \citenamefont {Zhou}, \citenamefont {Wang}, \citenamefont {Shan},\ and\
  \citenamefont {Xiao}}]{Chang2015}%
  \BibitemOpen
  \bibfield  {author} {\bibinfo {author} {\bibfnamefont {H.-R.}\ \bibnamefont
  {Chang}}, \bibinfo {author} {\bibfnamefont {J.}~\bibnamefont {Zhou}},
  \bibinfo {author} {\bibfnamefont {S.-X.}\ \bibnamefont {Wang}}, \bibinfo
  {author} {\bibfnamefont {W.-Y.}\ \bibnamefont {Shan}}, \ and\ \bibinfo
  {author} {\bibfnamefont {D.}~\bibnamefont {Xiao}},\ }\href {\doibase
  10.1103/PhysRevB.92.241103} {\bibfield  {journal} {\bibinfo  {journal} {Phys.
  Rev. B}\ }\textbf {\bibinfo {volume} {92}},\ \bibinfo {pages} {241103}
  (\bibinfo {year} {2015})}\BibitemShut {NoStop}%
\bibitem [{\citenamefont {Mastrogiuseppe}\ \emph {et~al.}(2016)\citenamefont
  {Mastrogiuseppe}, \citenamefont {Sandler},\ and\ \citenamefont
  {Ulloa}}]{Ulloa2016}%
  \BibitemOpen
  \bibfield  {author} {\bibinfo {author} {\bibfnamefont {D.}~\bibnamefont
  {Mastrogiuseppe}}, \bibinfo {author} {\bibfnamefont {N.}~\bibnamefont
  {Sandler}}, \ and\ \bibinfo {author} {\bibfnamefont {S.~E.}\ \bibnamefont
  {Ulloa}},\ }\href {\doibase 10.1103/PhysRevB.93.094433} {\bibfield  {journal}
  {\bibinfo  {journal} {Phys. Rev. B}\ }\textbf {\bibinfo {volume} {93}},\
  \bibinfo {pages} {094433} (\bibinfo {year} {2016})}\BibitemShut {NoStop}%
\bibitem [{\citenamefont {Kanazawa}\ and\ \citenamefont
  {Uchino}(2016)}]{Shun2016}%
  \BibitemOpen
  \bibfield  {author} {\bibinfo {author} {\bibfnamefont {T.}~\bibnamefont
  {Kanazawa}}\ and\ \bibinfo {author} {\bibfnamefont {S.}~\bibnamefont
  {Uchino}},\ }\href {\doibase 10.1103/PhysRevD.94.114005} {\bibfield
  {journal} {\bibinfo  {journal} {Phys. Rev. D}\ }\textbf {\bibinfo {volume}
  {94}},\ \bibinfo {pages} {114005} (\bibinfo {year} {2016})}\BibitemShut
  {NoStop}%
\bibitem [{\citenamefont {Zheng}\ \emph {et~al.}(2016)\citenamefont {Zheng},
  \citenamefont {Wang}, \citenamefont {Zhong},\ and\ \citenamefont
  {Duan}}]{Zheng2016}%
  \BibitemOpen
  \bibfield  {author} {\bibinfo {author} {\bibfnamefont {S.-H.}\ \bibnamefont
  {Zheng}}, \bibinfo {author} {\bibfnamefont {R.-Q.}\ \bibnamefont {Wang}},
  \bibinfo {author} {\bibfnamefont {M.}~\bibnamefont {Zhong}}, \ and\ \bibinfo
  {author} {\bibfnamefont {H.-J.}\ \bibnamefont {Duan}},\ }\href
  {https://www.nature.com/articles/srep36106?WT.feed_name=subjects_physical-sciences}
  {\bibfield  {journal} {\bibinfo  {journal} {Scientific reports}\ }\textbf
  {\bibinfo {volume} {6}},\ \bibinfo {pages} {36106} (\bibinfo {year}
  {2016})}\BibitemShut {NoStop}%
\bibitem [{\citenamefont {Feng}\ \emph {et~al.}(2010)\citenamefont {Feng},
  \citenamefont {Chen}, \citenamefont {Gao}, \citenamefont {Wang},\ and\
  \citenamefont {Zhang}}]{Feng2010}%
  \BibitemOpen
  \bibfield  {author} {\bibinfo {author} {\bibfnamefont {X.-Y.}\ \bibnamefont
  {Feng}}, \bibinfo {author} {\bibfnamefont {W.-Q.}\ \bibnamefont {Chen}},
  \bibinfo {author} {\bibfnamefont {J.-H.}\ \bibnamefont {Gao}}, \bibinfo
  {author} {\bibfnamefont {Q.-H.}\ \bibnamefont {Wang}}, \ and\ \bibinfo
  {author} {\bibfnamefont {F.-C.}\ \bibnamefont {Zhang}},\ }\href {\doibase
  10.1103/PhysRevB.81.235411} {\bibfield  {journal} {\bibinfo  {journal} {Phys.
  Rev. B}\ }\textbf {\bibinfo {volume} {81}},\ \bibinfo {pages} {235411}
  (\bibinfo {year} {2010})}\BibitemShut {NoStop}%
\bibitem [{\citenamefont {Shirakawa}\ and\ \citenamefont
  {Yunoki}(2014)}]{shirakawa2014}%
  \BibitemOpen
  \bibfield  {author} {\bibinfo {author} {\bibfnamefont {T.}~\bibnamefont
  {Shirakawa}}\ and\ \bibinfo {author} {\bibfnamefont {S.}~\bibnamefont
  {Yunoki}},\ }\href {\doibase 10.1103/PhysRevB.90.195109} {\bibfield
  {journal} {\bibinfo  {journal} {Phys. Rev. B}\ }\textbf {\bibinfo {volume}
  {90}},\ \bibinfo {pages} {195109} (\bibinfo {year} {2014})}\BibitemShut
  {NoStop}%
\bibitem [{\citenamefont {Liu}\ \emph {et~al.}(2009)\citenamefont {Liu},
  \citenamefont {Liu}, \citenamefont {Xu}, \citenamefont {Qi},\ and\
  \citenamefont {Zhang}}]{Liu2009}%
  \BibitemOpen
  \bibfield  {author} {\bibinfo {author} {\bibfnamefont {Q.}~\bibnamefont
  {Liu}}, \bibinfo {author} {\bibfnamefont {C.-X.}\ \bibnamefont {Liu}},
  \bibinfo {author} {\bibfnamefont {C.}~\bibnamefont {Xu}}, \bibinfo {author}
  {\bibfnamefont {X.-L.}\ \bibnamefont {Qi}}, \ and\ \bibinfo {author}
  {\bibfnamefont {S.-C.}\ \bibnamefont {Zhang}},\ }\href {\doibase
  10.1103/PhysRevLett.102.156603} {\bibfield  {journal} {\bibinfo  {journal}
  {Phys. Rev. Lett.}\ }\textbf {\bibinfo {volume} {102}},\ \bibinfo {pages}
  {156603} (\bibinfo {year} {2009})}\BibitemShut {NoStop}%
\bibitem [{\citenamefont {Sun}\ \emph {et~al.}(2015{\natexlab{b}})\citenamefont
  {Sun}, \citenamefont {Xu}, \citenamefont {Zhang},\ and\ \citenamefont
  {Zhou}}]{Jinhua2015}%
  \BibitemOpen
  \bibfield  {author} {\bibinfo {author} {\bibfnamefont {J.-H.}\ \bibnamefont
  {Sun}}, \bibinfo {author} {\bibfnamefont {D.-H.}\ \bibnamefont {Xu}},
  \bibinfo {author} {\bibfnamefont {F.-C.}\ \bibnamefont {Zhang}}, \ and\
  \bibinfo {author} {\bibfnamefont {Y.}~\bibnamefont {Zhou}},\ }\href {\doibase
  10.1103/PhysRevB.92.195124} {\bibfield  {journal} {\bibinfo  {journal} {Phys.
  Rev. B}\ }\textbf {\bibinfo {volume} {92}},\ \bibinfo {pages} {195124}
  (\bibinfo {year} {2015}{\natexlab{b}})}\BibitemShut {NoStop}%
\bibitem [{\citenamefont {Varma}\ and\ \citenamefont
  {Yafet}(1976)}]{Varma1976}%
  \BibitemOpen
  \bibfield  {author} {\bibinfo {author} {\bibfnamefont {C.~M.}\ \bibnamefont
  {Varma}}\ and\ \bibinfo {author} {\bibfnamefont {Y.}~\bibnamefont {Yafet}},\
  }\href {\doibase 10.1103/PhysRevB.13.2950} {\bibfield  {journal} {\bibinfo
  {journal} {Phys. Rev. B}\ }\textbf {\bibinfo {volume} {13}},\ \bibinfo
  {pages} {2950} (\bibinfo {year} {1976})}\BibitemShut {NoStop}%
\bibitem [{\citenamefont {Aji}\ \emph {et~al.}(2008)\citenamefont {Aji},
  \citenamefont {Varma},\ and\ \citenamefont {Vekhter}}]{Aji2008}%
  \BibitemOpen
  \bibfield  {author} {\bibinfo {author} {\bibfnamefont {V.}~\bibnamefont
  {Aji}}, \bibinfo {author} {\bibfnamefont {C.~M.}\ \bibnamefont {Varma}}, \
  and\ \bibinfo {author} {\bibfnamefont {I.}~\bibnamefont {Vekhter}},\ }\href
  {\doibase 10.1103/PhysRevB.77.224426} {\bibfield  {journal} {\bibinfo
  {journal} {Phys. Rev. B}\ }\textbf {\bibinfo {volume} {77}},\ \bibinfo
  {pages} {224426} (\bibinfo {year} {2008})}\BibitemShut {NoStop}%
\bibitem [{\citenamefont {Ma}\ \emph {et~al.}(2018)\citenamefont {Ma},
  \citenamefont {Chen}, \citenamefont {Liu},\ and\ \citenamefont
  {Xie}}]{Ma2018}%
  \BibitemOpen
  \bibfield  {author} {\bibinfo {author} {\bibfnamefont {D.}~\bibnamefont
  {Ma}}, \bibinfo {author} {\bibfnamefont {H.}~\bibnamefont {Chen}}, \bibinfo
  {author} {\bibfnamefont {H.}~\bibnamefont {Liu}}, \ and\ \bibinfo {author}
  {\bibfnamefont {X.~C.}\ \bibnamefont {Xie}},\ }\href {\doibase
  10.1103/PhysRevB.97.045148} {\bibfield  {journal} {\bibinfo  {journal} {Phys.
  Rev. B}\ }\textbf {\bibinfo {volume} {97}},\ \bibinfo {pages} {045148}
  (\bibinfo {year} {2018})}\BibitemShut {NoStop}%
\bibitem [{\citenamefont {L\"u}\ \emph {et~al.}(2019)\citenamefont {L\"u},
  \citenamefont {Deng}, \citenamefont {Ke}, \citenamefont {Guo},\ and\
  \citenamefont {Zhang}}]{Lv2019}%
  \BibitemOpen
  \bibfield  {author} {\bibinfo {author} {\bibfnamefont {H.-F.}\ \bibnamefont
  {L\"u}}, \bibinfo {author} {\bibfnamefont {Y.-H.}\ \bibnamefont {Deng}},
  \bibinfo {author} {\bibfnamefont {S.-S.}\ \bibnamefont {Ke}}, \bibinfo
  {author} {\bibfnamefont {Y.}~\bibnamefont {Guo}}, \ and\ \bibinfo {author}
  {\bibfnamefont {H.-W.}\ \bibnamefont {Zhang}},\ }\href {\doibase
  10.1103/PhysRevB.99.115109} {\bibfield  {journal} {\bibinfo  {journal} {Phys.
  Rev. B}\ }\textbf {\bibinfo {volume} {99}},\ \bibinfo {pages} {115109}
  (\bibinfo {year} {2019})}\BibitemShut {NoStop}%
\bibitem [{\citenamefont {Sun}\ \emph {et~al.}(2018)\citenamefont {Sun},
  \citenamefont {Wang}, \citenamefont {Hu}, \citenamefont {Li},\ and\
  \citenamefont {Xu}}]{Jinhua2018}%
  \BibitemOpen
  \bibfield  {author} {\bibinfo {author} {\bibfnamefont {J.-H.}\ \bibnamefont
  {Sun}}, \bibinfo {author} {\bibfnamefont {L.-J.}\ \bibnamefont {Wang}},
  \bibinfo {author} {\bibfnamefont {X.-T.}\ \bibnamefont {Hu}}, \bibinfo
  {author} {\bibfnamefont {L.}~\bibnamefont {Li}}, \ and\ \bibinfo {author}
  {\bibfnamefont {D.-H.}\ \bibnamefont {Xu}},\ }\href {\doibase
  10.1103/PhysRevB.97.035130} {\bibfield  {journal} {\bibinfo  {journal} {Phys.
  Rev. B}\ }\textbf {\bibinfo {volume} {97}},\ \bibinfo {pages} {035130}
  (\bibinfo {year} {2018})}\BibitemShut {NoStop}%
\bibitem [{\citenamefont {Deng}\ \emph {et~al.}(2018)\citenamefont {Deng},
  \citenamefont {L{\"u}}, \citenamefont {Ke}, \citenamefont {Guo},\ and\
  \citenamefont {Zhang}}]{deng2018}%
  \BibitemOpen
  \bibfield  {author} {\bibinfo {author} {\bibfnamefont {Y.-H.}\ \bibnamefont
  {Deng}}, \bibinfo {author} {\bibfnamefont {H.-F.}\ \bibnamefont {L{\"u}}},
  \bibinfo {author} {\bibfnamefont {S.-S.}\ \bibnamefont {Ke}}, \bibinfo
  {author} {\bibfnamefont {Y.}~\bibnamefont {Guo}}, \ and\ \bibinfo {author}
  {\bibfnamefont {H.-W.}\ \bibnamefont {Zhang}},\ }\href
  {https://iopscience.iop.org/article/10.1088/1361-648X/aae21d/meta} {\bibfield
   {journal} {\bibinfo  {journal} {Journal of Physics: Condensed Matter}\
  }\textbf {\bibinfo {volume} {30}},\ \bibinfo {pages} {435602} (\bibinfo
  {year} {2018})}\BibitemShut {NoStop}%
\bibitem [{\citenamefont {Zubkov}\ and\ \citenamefont
  {Lewkowicz}(2018)}]{zubkov2018}%
  \BibitemOpen
  \bibfield  {author} {\bibinfo {author} {\bibfnamefont {M.}~\bibnamefont
  {Zubkov}}\ and\ \bibinfo {author} {\bibfnamefont {M.}~\bibnamefont
  {Lewkowicz}},\ }\href@noop {} {\bibfield  {journal} {\bibinfo  {journal}
  {Annals of Physics}\ }\textbf {\bibinfo {volume} {399}},\ \bibinfo {pages}
  {26} (\bibinfo {year} {2018})}\BibitemShut {NoStop}%
\bibitem [{\citenamefont {Zheng}\ and\ \citenamefont
  {Hasan}(2018)}]{Zheng2018}%
  \BibitemOpen
  \bibfield  {author} {\bibinfo {author} {\bibfnamefont {H.}~\bibnamefont
  {Zheng}}\ and\ \bibinfo {author} {\bibfnamefont {M.~Z.}\ \bibnamefont
  {Hasan}},\ }\href {\doibase 10.1080/23746149.2018.1466661} {\bibfield
  {journal} {\bibinfo  {journal} {Advances in Physics: X}\ }\textbf {\bibinfo
  {volume} {3}},\ \bibinfo {pages} {1466661} (\bibinfo {year}
  {2018})}\BibitemShut {NoStop}%
\end{thebibliography}%
	
\begin{appendix}

\newpage
\widetext \vspace{0.5cm}


\global\long\def\theequation{S\arabic{equation}}
\global\long\def\thefigure{S\arabic{figure}}  

\global\long\def\bibnumfmt#1{[S#1]}
\global\long\def\citenumfont#1{S#1}

\section{}

	The $4\times 4$ Hamiltonian of the type-II WSM $h_0(\mathbf{k})$ given in Eq. \ref{Eq:H0} can be easily diagonalized through 
\begin{equation}
\begin{aligned}
\mathcal{V}^{\dagger}h_{0}(\mathbf{k})\mathcal{V} = \mathcal{E}(\mathbf{k}). 
\end{aligned}
\end{equation}
$\mathcal{E}(\mathbf{k})$ is the diagonal matrix whose diagonal elements are the eigen-energies. 
The elements of the vector matrix $\mathcal{V}$ are given by	
\begin{equation}
\begin{aligned}
\Phi_{1i} &=   \frac{-b(t^2k_z^2+M_\mathbf{k}^2+q\nu_\mathbf{k})+(p\nu_\mathbf{k}+(-1)^{i+1}tk_zb)\sqrt{\eta_\mathbf{k}+2q\nu_\mathbf{k}}+tk_z(-T^2(k_x^2+k_y^2)-q\nu_\mathbf{k}-b^2)}{T(k_x+ik_y)(b-tk_z)M_\mathbf{k}} \cdot C_i,\\
\Phi_{2i} &= \frac{-q\nu_\mathbf{k}-t^2k_z^2+(-1)^{i+1}tk_z\sqrt{\eta_\mathbf{k}+2q\nu_\mathbf{k}}}{(b-tk_z)M_\mathbf{k}} \cdot C_i,\\
\Phi_{3i} &= \frac{-q\nu_\mathbf{k}-b^2+(-1)^{i+1}b\sqrt{\eta_\mathbf{k}+2q\nu_\mathbf{k}}}{T(k_x+ik_y)(b-tk_z)} \cdot C_i,\\
\Phi_{4i} &= C_i.\\
\end{aligned}
\end{equation}
$C_i$ $(i=1,2,3,4)$ are normalization factors, and $p$ and $q$ are simply numbers. When $i\in\{1,2\}$, $q=-1$, otherwise $q=+1$. When $i\in\{1,4\}$ $p=-1$, otherwise $p=+1$.
The eigenstates of the tilted Dirac cone is given by
\begin{equation}
\begin{aligned}
\Gamma_\mathbf{k}=\mathcal{V}^\dagger\Psi_\mathbf{k}.
\end{aligned}
\end{equation}
Where $\Psi_{\mathbf{k}}= \{a_{\mathbf{k}\uparrow},  a_{\mathbf{k}\downarrow}, b_{\mathbf{k}\uparrow}, b_{\mathbf{k}\downarrow}\}^T$, and  $\Gamma_\mathbf{k}=\{\gamma_{\mathbf{k}1}, \gamma_{\mathbf{k}2}, \gamma_{\mathbf{k}3}, \gamma_{\mathbf{k}4}\}^T$.
Then $H_0$ in its diagonal basis writes
\begin{equation}
\begin{aligned}
H_0 = \sum_\mathbf{k} h_0(\mathbf{k}) =  \sum_{\mathbf{k}i} \epsilon_{\mathbf{k}i}\gamma_{\mathbf{k}i}^\dagger \gamma_{\mathbf{k}i}, \ \  (i=1,2,3,4). \\
\end{aligned}
\end{equation}
We define a function which can be used to simplify the coordinate space spin-spin correlation function    
\begin{equation}\label{Eq:21}
\begin{aligned}
\mathcal{A}_{mn} (\mathbf{r}) = \sum_{\mathbf{k}i} \Phi_{mi}^* (\mathbf{k}) \chi_{in} (\mathbf{k})a_{\mathbf{k}i}e^{-i\mathbf{kr}}, 
\end{aligned}
\end{equation}
where the numbers $\{i,m,n\}=\{1,2,3,4\}$. 
Both the $a$ and $b$ orbits of the type-II WSM contribute to the spin-spin correlation between the magnetic impurity and the conduction electron located on $\mathbf{r}$. Subsequently, the correlation function consists of two parts,     
$J_{uv}(\mathbf{r})= \langle S_{a}^u (\mathbf{r})S_d^v(0)+S_{b}^u (\mathbf{r})S_d^v(0)\rangle=J_{uv}^a(\mathbf{r})+J_{uv}^b(\mathbf{r})$. Here $u,v=x,y,z$, and $\langle \cdots \rangle$ denotes the ground state average. 
The spin-spin correlation function between a magnetic impurity and the conduction electrons from $a$ and $b$ orbits are given by 
\begin{equation}\label{Eq:sscorr}
\begin{aligned}
\mathbf{J}_{zz}^a(\mathbf{r})&=-\frac{1}{4}\left(\left|\mathcal{A}_{11}\right|^2-\left|\mathcal{A}_{12}\right|^2-\left|\mathcal{A}_{21}\right|^2+\left|\mathcal{A}_{22}\right|^2\right),\\
\mathbf{J}_{zz}^b(\mathbf{r})&=-\frac{1}{4}\left(\left|\mathcal{A}_{31}\right|^2-\left|\mathcal{A}_{32}\right|^2-\left|\mathcal{A}_{41}\right|^2+\left|\mathcal{A}_{42}\right|^2\right),\\
\mathbf{J}_{xx}^a(\mathbf{r})&=-\frac{1}{2}\left[\mathrm{Re}\left(\mathcal{A}_{12}\mathcal{A}_{21}^*\right)+\mathrm{Re}\left(\mathcal{A}_{11}\mathcal{A}_{22}^*\right)\right],\\	
\mathbf{J}_{xx}^b(\mathbf{r})&=-\frac{1}{2}\left[\mathrm{Re}\left(\mathcal{A}_{32}\mathcal{A}_{41}^*\right)+\mathrm{Re}\left(\mathcal{A}_{31}\mathcal{A}_{42}^*\right)\right],\\		 \mathbf{J}_{yy}^a(\mathbf{r})&=-\frac{1}{2}\left[-\mathrm{Re}\left(\mathcal{A}_{12}\mathcal{A}_{21}^*\right)+\mathrm{Re}\left(\mathcal{A}_{11}\mathcal{A}_{22}^*\right)\right],\\	
\mathbf{J}_{yy}^b(\mathbf{r})&=-\frac{1}{2}\left[-\mathrm{Re}\left(\mathcal{A}_{32}\mathcal{A}_{41}^*\right)+\mathrm{Re}\left(\mathcal{A}_{31}\mathcal{A}_{42}^*\right)\right],\\			
\mathbf{J}_{xy}^a(\mathbf{r})&=\frac{1}{2}\left[\mathrm{Im}\left(\mathcal{A}_{12}^*\mathcal{A}_{21}\right)+\mathrm{Im}\left(\mathcal{A}_{11}\mathcal{A}_{22}^*\right)\right],\\	
\mathbf{J}_{xy}^b(\mathbf{r})&=\frac{1}{2}\left[	\mathrm{Im}\left(\mathcal{A}_{32}^*\mathcal{A}_{41}\right)+\mathrm{Im}\left(\mathcal{A}_{31}\mathcal{A}_{42}^*\right)\right],\\
\mathbf{J}_{xz}^a(\mathbf{r})&=-\frac{1}{2}\left[\mathrm{Re}\left(\mathcal{A}_{11}\mathcal{A}_{21}^*\right)-\mathrm{Re}\left(\mathcal{A}_{12}\mathcal{A}_{22}^*\right)\right],\\	
\mathbf{J}_{xz}^b(\mathbf{r})&=-\frac{1}{2}\left[\mathrm{Re}\left(\mathcal{A}_{31}\mathcal{A}_{41}^*\right)-\mathrm{Re}\left(\mathcal{A}_{32}\mathcal{A}_{42}^*\right)\right],\\	
\mathbf{J}_{yz}^a(\mathbf{r})&=\frac{1}{2}\left[\mathrm{Im}\left(\mathcal{A}_{11}^*\mathcal{A}_{21}\right)+\mathrm{Im}\left(\mathcal{A}_{12}\mathcal{A}_{22}^*\right)\right],\\	
\mathbf{J}_{yz}^b(\mathbf{r})&=\frac{1}{2}\left[\mathrm{Im}\left(\mathcal{A}_{31}^*\mathcal{A}_{41}\right)+\mathrm{Im}\left(\mathcal{A}_{32}\mathcal{A}_{42}^*\right)\right],\\
\mathbf{J}_{yx}^a(\mathbf{r})&=\frac{1}{2}\left[\mathrm{Im}\left(\mathcal{A}_{12}^*\mathcal{A}_{21}\right)-\mathrm{Im}\left(\mathcal{A}_{11}\mathcal{A}_{22}^*\right)\right],\\	
\mathbf{J}_{yx}^b(\mathbf{r})&=\frac{1}{2}\left[	\mathrm{Im}\left(\mathcal{A}_{32}^*\mathcal{A}_{41}\right)-\mathrm{Im}\left(\mathcal{A}_{31}\mathcal{A}_{42}^*\right)\right],\\
\mathbf{J}_{zx}^a(\mathbf{r})&=-\frac{1}{2}\left[\mathrm{Re}\left(\mathcal{A}_{12}\mathcal{A}_{11}^*\right)-\mathrm{Re}\left(\mathcal{A}_{22}\mathcal{A}_{21}^*\right)\right],\\	
\mathbf{J}_{zx}^b(\mathbf{r})&=-\frac{1}{2}\left[\mathrm{Re}\left(\mathcal{A}_{32}\mathcal{A}_{31}^*\right)-\mathrm{Re}\left(\mathcal{A}_{42}\mathcal{A}_{41}^*\right)\right],\\
\mathbf{J}_{zy}^a(\mathbf{r})&=\frac{1}{2}\left[\mathrm{Im}\left(\mathcal{A}_{11}\mathcal{A}_{12}^*\right)+\mathrm{Im}\left(\mathcal{A}_{22}\mathcal{A}_{21}^*\right)\right],\\
\mathbf{J}_{zy}^b(\mathbf{r})&=\frac{1}{2}\left[\mathrm{Im}\left(\mathcal{A}_{31}\mathcal{A}_{32}^*\right)+\mathrm{Im}\left(\mathcal{A}_{42}\mathcal{A}_{41}^*\right)\right].\\	
\end{aligned}
\end{equation}
\begin{figure*}[htb!]
	\begin{center}
		\includegraphics[scale=0.65, bb=280 120 550 400]{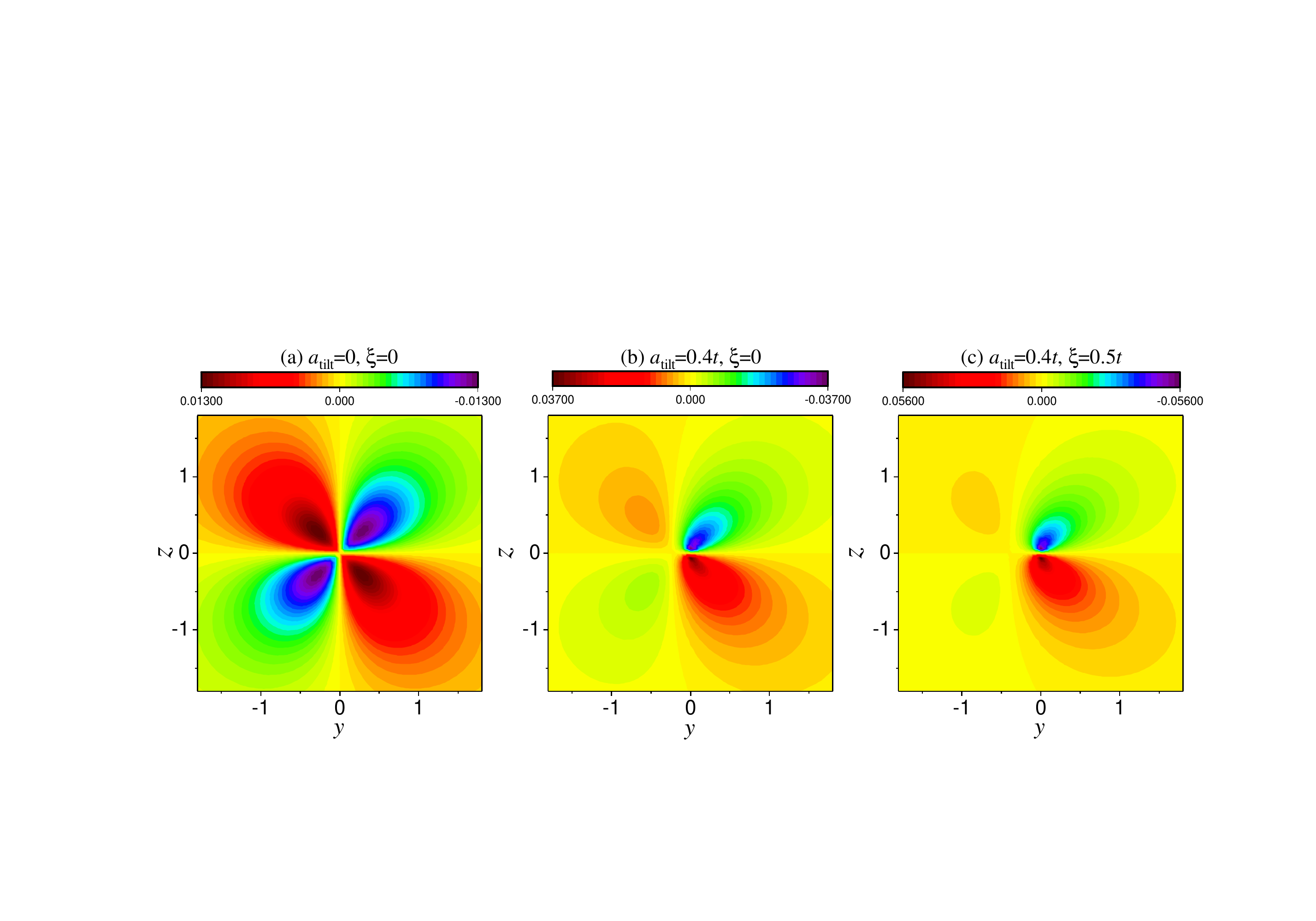}
	\end{center}
	\caption{(Color online). $J_{zy}(\mathbf{r})$ for three combinations of $a_{tilt}$ and $\xi$ on the $y$-$z$ coordinate plane for $b=0.5t$. The values are different from $J_{yz}(\mathbf{r})$ given in Fig. \ref{2_yz}, but the symmetry property is the same.    } \label{zy}
\end{figure*} 

$\mathcal{A}_{mn}(\mathbf{r})$ given in Eq. \ref{Eq:21} are complex numbers, so $J_{uv}(\mathbf{r})\neq J_{vu}(\mathbf{r})$ in general. 
Below we will mainly analyze the nonzero off-diagonal components of spin-spin correlation on the three principal planes. 

$J_{xy}(\mathbf{r})$ and $J_{yx}(\mathbf{r})$ are nonzero on the $x$-$y$ plane, and also on the $x$-$z$ plane in presence of tilting terms. We find that the second terms of $J_{xy}^a(\mathbf{r})$ and $J_{xy}^b(\mathbf{r})$ cancel with each other, meaning that $\mathrm{Im}(\mathcal{A}_{11}\mathcal{A}_{22}^*)+\mathrm{Im}(\mathcal{A}_{31}\mathcal{A}_{42}^*)=0$. Consequently, on the $x$-$y$ and $x$-$z$ coordinate planes, $J_{xy}(\mathbf{r})=J_{yx}(\mathbf{r})$. 

$J_{yz}(\mathbf{r})$ and $J_{zy}(\mathbf{r})$ are nonzero on $y$-$z$ plane, and also on the $x$-$z$ plane in the presence of tilting terms.
On the $y$-$z$ plane, $J_{yz}(\mathbf{r})\neq J_{zy}(\mathbf{r})$, and we plot the results of $J_{zy}(\mathbf{r})$ on the $y$-$z$ plane on Fig. \ref{zy}.

\begin{figure*}[htb!]
	\begin{center}
		\includegraphics[scale=0.65, bb=280 120 550 300]{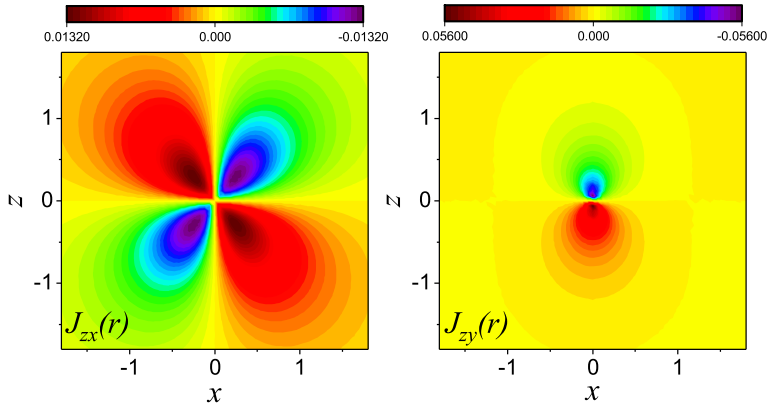}
	\end{center}
	\caption{(Color online). $J_{zx}(\mathbf{r})$ and $J_{zy}(\mathbf{r})$ on the $x$-$z$ plane for $b=0.5t$, $a_{tilt}=0.4t$ and $\xi=0.5t$.    } \label{zx}
\end{figure*}

 On the $x$-$z$ plane, and in the absence of $a_{tilt}$ and $\xi$, the model Hamiltonian of the type-II WSM preserves the rotational symmetry about the $z$ direction. Hence one may have $J_{xz}(x,z) = J_{yz}(y,z)$ and $J_{zx}(x,z)=J_{zy}(y,z)$. In Fig. \ref{zx} we show the results of non-zero off-diagonal components of spin-spin correlation function on the $x$-$z$ plane. Remarkably, we find that $J_{zy}(\mathbf{r})$ is negative while $z>0$ while $J_{yz}(\mathbf{r})$ is positive. $J_{zx}(\mathbf{r})$ is different in values in comparison with 
$J_{xz}(\mathbf{r})$ plotted in Fig. \ref{3_xz}. 

\end{appendix}

\end{document}